\documentclass[
reprint,
superscriptaddress,
nofootinbib,
amsmath,amssymb,
twocolumn, aps,
prd,
]{revtex4}

\usepackage[utf8]{inputenc}
\usepackage{amsmath}
\usepackage{mathtools}
\usepackage{amssymb,amstext,amsfonts}
\usepackage{graphicx}
\usepackage{float}
\usepackage{microtype}
\usepackage{subfigure}
\usepackage{multirow}
\usepackage{hyperref} 
\usepackage{lineno}
\usepackage{color}
\usepackage{comment}
\usepackage[normalem]{ulem}

\newcommand{\be}{ \begin{equation} }
\newcommand{\ee}{ \end{equation}}

\usepackage{amssymb}
\usepackage{pifont}
\newcommand{\cmark}{\ding{51}}%
\newcommand{\xmark}{\ding{55}}%

\begin{document}

\title{Enhancing modified gravity detection from gravitational-wave observations\\ using the Parametrized ringdown spin expansion coefficients formalism}

\author{Gregorio Carullo}
\affiliation{Dipartimento di Fisica ``Enrico Fermi'', Universit\`a di Pisa, Pisa I-56127, Italy}
\affiliation{INFN sezione di Pisa, Pisa I-56127, Italy}

\date{\today}

\begin{abstract}

Harvesting the full potential of black hole spectroscopy demands realising the importance of casting constraints on modified theories of gravity in a framework as general and robust as possible.
Requiring more stringent -- yet well-motivated -- beyond General Relativity (GR) parametrizations improves the inference drawn from available GW data, substantially decreasing the errors on deviation parameters. This implies a reduction in the number of signals needed to detect a deviation from GR predictions and an increase of the number of GR-violating coefficients that can be meaningfully constrained with a given number of signals. To this end, we apply to LIGO-Virgo observations a high-spin version of the Parametrized ringdown spin expansion coefficients (ParSpec) formalism, encompassing large classes of modified theories of gravity. We constrain the lowest-order perturbative deviation of the fundamental ringdown frequency to be $\delta\omega^{0}_{220} = {-0.05}^{+0.05}_{-0.05}$, when assuming adimensional beyond-GR couplings, substantially improving upon previously published results. We also establish upper bounds $\ell_{p=2} < 23 \, \mathrm{km}$, $\ell_{p=4} < 35 \, \mathrm{km}$, $\ell_{p=6} < 42 \, \mathrm{km}$ on the scale $\ell_p$ at which the appearance of new physics is disfavoured, depending on the mass dimension $p$ of the ringdown coupling. These bounds exceed the ones obtained by previous analyses or are competitive with existing ones, depending on the specific alternative theory considered, and promise to quickly improve as the number of detectors, sensitivity and duty-cycle of the gravitational-wave network steadily increases.

\end{abstract}

\maketitle


\section{Introduction} 
Within the framework of GR and the current Standard Model of particle physics, black holes (BHs) are the inescapable final state of massive objects ($\gtrsim 3 M_{\odot}$) gravitational collapse. Investigating the structure of extremely compact objects, confirming their BH nature and testing the predictions of GR in the BH near-horizon regime, is one of the main scientific goals of gravitational-wave astronomy and BH imaging~\cite{Aasi:2013wya, O3a_TGR, Akiyama:2019fyp, Psaltis:2018xkc, Akiyama:2019cqa}. Among the wide variety of BH characteristic signatures, a considerable amount of effort has been devoted to bound~\cite{TGR-LVC2016, Brito:2018rfr, CARULLO-DELPOZZO-VEITCH, ISI-GIESLER-FARR-SCHEEL-TEUKOLSKY, O3a_TGR, Laghi:2020rgl} deviations in the spectrum dominating the ringdown process, the GW emission process driving the post-merger phase of a binary black hole (BBH) coalescence. Ringdown emission is sensitive to the space-time structure close to the last stable photon orbit~\cite{PhysRevD.31.290, Cardoso:2016rao}, allowing precision tests of gravity in the strong gravitational-field regime. Deviations from the predictions of BH dynamics in GR could either be caused by a modification to the theory action~\cite{Psaltis:2008bb} or by exotic compact objects (ECOs) mimicking BHs behaviour. The latter class of objects could be constituted by new species of particles or by unknown stable solution of Einstein's field equations arising from underlying quantum effects~\cite{Cardoso:2019rvt}.\\

The latest catalog of observations~\cite{O3a_catalog} announced by the LIGO-Virgo collaborations~\cite{AdvLIGO,AdvVirgo} has brought the total number of confirmed gravitational-wave events to 50. The increasing amount of detections is shifting the focus of gravitational-wave observational science from single events properties to population properties. The signal-to-noise ratio (SNR) of these observations is weak compared to the ones that should be achieved by both ground-based or satellite future experiments ~\cite{ET, CE, LISA}, but the number of detected events by the current gravitational-waves detectors network will soon grow to hundreds per year~\cite{Aasi:2013wya}. Although most of these signals will have a small SNR, thus providing weak constraints per-se, bounds on beyond-GR parametrizations will steadily improve with the number of additional detections~\cite{Carullo:2018sfu, Brito:2018rfr, Bhagwat:2019dtm, Bhagwat:2019bwv}. Current detector network capability of testing the BH ringdown emission could be further enhanced by cavity detuning techniques. Applied in coordination with LISA early-inspiral observations, detuning would allow ground-based detectors to vastly increase their sensitivity around the expected ringdown frequency band and boost spectroscopy measurements precision~\cite{Tso:2018pdv}. Such a coordination strategy highlights the new and exciting possibilities that will be opened by multi-band GW observations, enhancing the achievable precision on tests of GR obtainable by single class of detectors alone~\cite{Carson:2019rda, Carson:2019kkh, Datta:2020vcj, Gupta:2020lxa}.\\

Given the wide range of possible modifications from GR predictions, when performing the aforementioned tests it is important to use generic and robust parametrizations.
A choice extensively employed in the literature, e.g. by the LIGO-Virgo collaborations~\cite{O3a_TGR}, allows to directly cast constraints on observable quantities by considering parametrized deviations to ringdown frequencies and damping times. Such an alternative has the advantage of capturing deviations that arise in the observed quantities of interest, without being dependent on a specific alternative theory of gravity. 
While being quite general, this approach has the disadvantage of being source dependent, since deviation parameters will also depend on the mass and spin of the remnant BH and thus generally differ for different astrophysical sources. 
Although it is still possible to combine multiple events, construct a global figure of merit on the agreement of GR with observations (Bayes Factor) and estimate the population distribution of putative deviations~\cite{Hierarchical_I, Hierarchical_II, O3a_TGR}, an approach where this additional step would not be needed could decrease the number of signals needed to be considered in order to claim the detection of a deviation.\\
Another, sometimes unappreciated, point to be considered when discussing such tests is that in presence of a GR violation not only the parametrized coefficients, but also the physical BH mass and spin would also differ from their GR predicted value. These two additional intrinsic parameters\footnote{The full set of parameters characterising the GW signal independently of the position of the observer in the sky, such as BH masses and spins or GR deviation coefficients are referred to as \textit{intrinsic parameters.}} need to be marginalised over, greatly reducing the resolving power of the analysis. Such a reduction stems from the possibly strong correlations in the multi-dimensional space of intrinsic parameters. Detecting a deviation from GR, consequently requires an SNR (or number of signals) larger than expected from uncertainty estimates solely based on considering GR deviation parameters.\\
A formalism not suffering from the aforementioned resolving-power reduction and allowing to extract source-independent GR deviation parameters, has been recently proposed in Ref.~\cite{ParSpec}. The Parametrized ringdown spin expansion coefficients (ParSpec) formalism, described in detail in the next sections and here applied for the first time to observational data, provides a robust and accurate parametrization covering, at the perturbative level, large classes of modified gravity theories, such as Effective Field Theories of beyong-GR gravity, Einstein-scalar-Gauss-Bonnet or dynamical Chern-Symons gravity. Current measurement errors require to extend the validity of the numerical fits originally presented in Ref.~\cite{ParSpec} to higher values of the BH spin parameter, an extension easily performed being the formalism fully capable to accommodate for arbitrarily high values of the BH spin.
In this manuscript, we will apply this high-spin version of the ParSpec formalism to current LVC observations, showing how it allows to: i) place tighter constraints on BH spectra deviation parameters with respect to the ones present in the literature, hence accelerating the detection of possible deviations from the predictions of GR; ii) place competitive bounds on the energy scale at which effects due to new physics (including corrections from Einstein-scalar-Gauss-Bonnet, dynamical Chern-Simons or Effective Field Theories terms in the action) can manifest in BBH systems.\\

The paper is organised as follows: in Sec.~\ref{sec:ParSpec_basic} we review the ParSpec formalism, highlighting the cases of interest for our analysis. In Sec.~\ref{sec:ParSpec_highspin} we include in the expansion higher order terms to accommodate for high spins, an important step in order to perform an analysis on LIGO-Virgo data. The main section of the paper, Sec.~\ref{sec:Observations}, deals with the presentation and discussion of results obtained by applying the ParSpec framework to merger-ringdown observations from the LIGO-Virgo collaborations, comparing them to previously known bounds. Conclusions and future prospects are presented in Sec.~\ref{sec:Conclusions}. Throughout the manuscript both $c=G=1$ units, together with median and $90\%$ credible levels (CL) for parameters point estimates are used, except where explicitly stated otherwise.

\section{ParSpec}
\label{sec:ParSpec_basic}

\subsection{Basic formalism}
In this subsection we will review the ParSpec formalism, highlighting the features especially relevant to data analysis, and the classes of beyond-GR theories that can be mapped to this framework. For a complete account, see Ref.~\cite{ParSpec}.\\
Perturbed BHs relax towards their equilibrium state by emitting GWs. The main contribution to the GW signal, at intermediate times, is due to a superposition of damped normal modes, also called quasi-normal modes (QNMs), exhibiting a discrete complex spectrum that can be computed from BH perturbation theory (for a review, see Ref.~\cite{QNM_review_BCS}).
This prediction allows to test for the presence of modifications to the GR QNM emission from perturbed BHs "ringing down" towards their stationary state.\\
The ParSpec framework, aimed at implementing a suitable parametrisation for such a test, can be constructed starting from considering the following parametrization:

\be\label{eq:generic_expansion}
    \begin{aligned}
        \omega &= \omega^{Kerr} \cdot (1 + \delta\omega )\, ,\\
        \tau &= \tau^{Kerr} \cdot (1 + \delta\tau) \, .
    \end{aligned}
\ee 

valid for each QNM, where $\omega$ ($\tau$) is the BH QNM frequency (damping time) in a generic beyond-GR theory, while $\omega^{Kerr}$ ($\tau^{Kerr}$) represent the GR values as predicted by the Kerr solution. The parameters $\delta\omega, \delta\tau$ represent parametric deviations from the Kerr predictions.
Ignoring sources of systematic errors that will be discussed in the next sections, these deviation parameters can take non-zero values due to environmental effects, extra-charges beside the BH mass and spin, an ECO mimicking the BH emission or modifications to the BH GR-dynamics due to an alternative theory of gravity.
As already discussed, although the expansion presented in Eqs.~\ref{eq:generic_expansion} is generally valid, a drawback arises from the fact that the deviation parameters will depend on the specific source parameters (BH mass and spin) and thus deviation values will differ for different sources. Defining source-independent parameters would allow to directly combine constraints from multiple observations. The parametrization efficiency could also be improved (in the sense that a smaller number of observations is needed to confidently detect a deviation) by imposing a specific dependence on the BH mass and spin.
A way to realise these two desiderata is to consider a perturbative setting, through the following bivariate expansion, valid for a given GW source:

\be\label{eq:ParSpec_expansion}
    \begin{aligned}
        \omega_K &= \frac{1}{M} \, \sum_{j=0}^{N_{max}} \, \chi^j \, \omega^{(j)}_K \, (1+\gamma \, \delta \omega_K^{(j)})\, ,\\
        \tau_K &= M \, \sum_{j=0}^{N_{max}} \, \chi^j \, \tau^{(j)}_K \, (1+\gamma \, \delta \tau_K^{(j)}) \, .
        \end{aligned}
\ee
where $K$ labels the QNM considered, $\omega_K^{(j)}$, $\tau_K^{(j)}$ are the source-independent numerical coefficients governing the spin expansion in GR, $M, \chi$ represent the \textit{GR values} of the BH mass and spin and $N_{max}$ is the maximum expansion order considered. 
The specific scenario giving rise to a modification in the GR expansion is encoded into the $\delta \omega_K^{(j)}$, $\delta \tau_K^{(j)}$ coefficients and in the $\gamma$ parameter. The $\delta \omega_K^{(j)}$, $\delta \tau_K^{(j)}$ coefficients set the spin-expansion structure of the modification under consideration, and are thus numerical constants. The parameter $\gamma$ is a (possibly source-dependent) coupling constant, discussed below. Since a modification in the BH dynamics will modify not only the spectrum, but also the BH intrinsic parameters, the expansion should in principle be written in terms of non-GR BH mass and spin parameters $\bar{M}, \bar{\chi}$. Nonetheless, as proven in Ref.~\cite{ParSpec}, at the perturbative level considered, modifications to the BH mass and spin can be absorbed in the deviation parameters already defined \footnote{This implies that the deviation parameters considered will include a combination of intrinsic corrections to the spectrum and corrections to the mass and spin of the remnant BH, as discussed in Appendix A of Ref.~\cite{ParSpec}. At the perturbative level, these two types of corrections cannot be disentangled.}. Hence, the BH parameters measured by assuming that GR is the correct description of the BH dynamics ($M, \chi$), can be used in computing the spectrum. This simple observation will have a great impact on the analysis, as discussed in the next sections. The perturbative assumption, justified by the absence of any macroscopic deviation in LVC observations, is imposed by requiring $\gamma \, \delta\omega_K^{(j)} \ll 1$ ($\gamma \, \delta\tau_K^{(j)} \ll 1$) for all values of $\{K,j\}$ considered. The adimensional parameter $\gamma$ can be expressed in terms of a (possibly dimensionful) coupling constant $\alpha$. If we denote by $p$ the mass dimension of $\alpha$, we can express $\gamma$ in terms of the BH mass $M_s$ in the source frame, which sets the energy scale of the process under consideration:

\be
    \gamma = \frac{\alpha}{M_s^p} = \frac{\alpha \,(1+z)^p}{M^p} := \left(\frac{\ell \, c^2 \, (1+z)}{G\, M}\right)^p \,, 
\ee

where we expressed $\alpha$ in terms of a parameter $\ell$, representing the typical length scale below which corrections from new physics affect the system. Since the spin-dependence of the spectrum was extracted through the perturbative expansion, while the mass dependence has been explicitly factorised in the $\gamma$ parameter, $\alpha$ does not depend on the BH mass and spin. The $\alpha$ parameter is a dimensionful coupling for which two cases have to be distinguished. In the first case $\alpha$ is a source-independent parameter, function of the coupling constant(s) appearing in the action of the new theory and it is referred to as \textit{secondary hair}. The adimensional case ($p=0$) includes certain scalar-tensor theories, Einstein-Aether and Horava gravity~\cite{Berti:2015itd}, the $p=4$ case encompasses Einstein-scalar-Gauss-Bonnet and dynamical Chern-Simons gravity~\cite{Collodel:2019kkx, Alexander:2009tp, PhysRevD.91.069905, PhysRevD.79.084043, Maselli_2017, PhysRevD.84.087501, PhysRevD.92.083014, PhysRevD.83.104002, PhysRevD.100.104061, Okounkova:2017yby, Okounkova:2019dfo, Okounkova:2019zjf, Cano:2020cao}, while the $p=6$ case includes the (quite general) class of Effective Field Theories considered in Refs.~\cite{Endlich:2017tqa, PhysRevLett.121.251105, Sennett:2019bpc}. For an Effective Field Theory approach to scalar-tensor theories see instead Ref.~\cite{Franciolini:2018uyq, Noller:2019chl}.
In the second case, $\alpha$ is an additional BH charge independent from the mass and spin of the BH, a \textit{primary hair}. The presence of a BH $U(1)$ charge (which may correspond to a putative BH electric charge) is included in this latter case when $p=2$. In this scenario, the ParSpec formalism parametrizes violations of the Kerr hypothesis in the gravitational modes which are continuously connected to the $l = m = 2$ QNM in the Schwarzschild case~\cite{Kokkotas_charge, Berti:2005eb, PhysRevD.88.064048, Mark:2014aja, Zilhao:2012gp, PhysRevLett.114.151101, 2016JCAP...05..054C, Bozzola:2020mjx}. See Ref.~\cite{Cardoso:2019mqo, McManus:2019ulj, ParSpec} for an account on the classes of theories that can be mapped to the ParSpec formalism and Refs.~\cite{Berti:2015itd, Yagi:2016jml, Yunes:2013dva} for extensive reviews on modified theories of gravity and their observational effects. The redshift $z$ can be obtained from the luminosity distance assuming a standard cosmology (see Ref.~\cite{ParSpec} for a discussion on the validity of this assumption).
Without listing in detail all the implications that a measurements of the parameters considered in the ParSpec formalism entail (for a comprehensive review, see Ref.~\cite{Yunes:2016jcc}), we simply remark that constraints on these classes of theories have implications on the deviations from some of the theoretical pillars of GR, such as the Equivalence Principle and Lorentz-invariance. Furthermore, a variety of beyond-GR emission mechanisms, e.g. scalar fields emission, can lead a measurable effect in the observations considered and thus be constrained with BBH observations.
The outlined approach is perturbative in spirit and thus retains the same merger-ringdown structure as the one present in GR. Nonetheless, deviations parameters will peak away from zero even if large deviations are present or if the structure of the merger-ringdown differs from the one of GR (e.g. due to additional modes or non-exponential time-dependent terms, as in Ref.~\cite{Okounkova:2019dfo}). In this latter case the values of the deviation parameter can only be subject to a phenomenological interpretation and should be considered as detection indicators.\\

\subsection{Comparison to alternative parametrizations}

Given the extreme variety of new physics effects that could manifest in BBH coalescences (either in the form of alternative compact objects or stemming from a high-energy extension of GR), different choices to parametrize possible deviation from the GR BBH hypothesis are possible. In the following subsection, we compare the parametrization introduced by the ParSpec formalism to two notable alternatives, namely parametrizations of the BH metric and the parametrization employed in LVC catalogs. We discuss the advantages and drawbacks of the different choices in terms of observational analyses, arguing that the ParSpec formalism offers several advantages compared to common alternatives. These advantages are essentially due to the introduction of source-independent parameters and to a perturbative expansion allowing to reabsorb corrections of the mass and spin parameters into frequency deviation parameters, while maintaing a general applicability to large classes of possible deviations.\\

\subsubsection{Metric parametrizations}

An appealing possibility would be to consider an effective metric in a suitable coordinate frame, encompassing both the GR and beyond-GR case, to describe the BH spacetime. For a thorough overview of the properties and regions of validity of several modified Kerr metrics, see Ref.~\cite{Johannsen:2013rqa}. Considering deviations in the metric allows to cast constraints directly on a fundamental, although coordinate dependent, object determining the BH geometry. This approach was adopted in Ref.~\cite{PhysRevLett.125.141104}, where observations of the M87 BH shadow obtained from the Einstein Horizon Telescope (EHT), were employed to constrain the second post-Newtonian (2 PN) terms of effective non-Kerr metrics (but see Ref.~\cite{Volkel:2020xlc} for discussions about the underlying statistical assumptions and Ref.~\cite{PhysRevD.103.024023} for a critical assessment of all the underlying astrophysical assumptions entering the analysis). An extension combining shadow measurements with constraints from BBH inspiral signals was recently considered in Ref.~\cite{Psaltis:2020ctj}, finding no deviations from the predictions put forward by the Kerr metric in GR.
The same approach could in principle be used to extract constraints from the merger-ringdown regime of BBH remnants on an effective metric. This possibility was explicitly considered in the case of non-spinning BHs in Ref.~\cite{Volkel:2020daa}, for a simulated dataset in the context of next-generation GW detectors. Unfortunately, extending this procedure to the rotating case, while maintaining an agnostic approach, presents non-trivial complications~\cite{Volkel:2020daa}. A possibility to overcome some of these complications in the rotating case would be to consider an eikonal approximation, as described in Ref.~\cite{Glampedakis:2017dvb} (for applications of the eikonal approximation to the spherically symmetric case of alternative theories of gravity, see Refs.~\cite{Glampedakis:2019dqh, Silva:2019scu}).
Although coming with its own set of approximations and thus not generally valid, this approach would constitute quite a robust and generic formalism to project the constraints obtained from observations of ringdown spectra from rotating BHs for the fundamental $\ell=|m|$ subset of modes. 
From the pure point of view of data analysis, constructing templates (or even simple numerical interpolants) of QNM frequencies as a function of an effective metric parameters would constitute an efficient way to test gravity in this approach. Such a scenario seems challenging from the theoretical point of view, but not unfeasible in principle once a suitable formulation of the perturbation equations is found. The construction of such a template would be an extremely interesting avenue to be explored in the near future. Instead, solving numerically the perturbation equations for a given value of the effective metric parameter(s) during the analysis itself currently seems unfeasible. In fact, the computational cost that such an operation would add to the exploration of the parameter space, at least within current state-of-the-art stochastic samplers, would presumably not be sustainable. Machine learning methods would likely need to be introduced in order to decrease computational cost. The most serious limitation of the aforementioned approaches focusing on parametrized metrics stems from the far richer dynamics that is present in beyond-GR theories when modifications to the field equations are consistently included~\cite{McManus:2019ulj}.\\

\subsubsection{LVC parametrization}

A set of analyses similar in spirit to the ones considered in this work was notably put forward by the LIGO-Virgo collaborations in the recent catalog of tests of General Relativity~\cite{O3a_TGR}. In the latter reference, a battery of tests was considered, investigating: possible modifications in both the generation and propagation of gravitational-waves, the nature of the coalescing objects, additional polarisations absent in GR and residuals in the detector strain after subtracting a best-fit waveform. None of these tests signalled a discrepancy among the predictions put forward by GR and the observed BBH population.\\ 
Of particular interest for this work are the set of tests exploring \textit{parametrized deviations} in the GR predicted waveform. 
The reason why presumably these tests are not formulated as perturbative expansions in terms of a "small" parameter is that this approach would narrow down the set of possible theories to which they are sensitive to. When considering a full inspiral-plunge-merger-ringdown analysis, in the inspiral regime the deviations are applied to a set of coefficients derived from PN theory in both the EOB and Phenom set of waveforms~\cite{Bohe:2016gbl, Schmidt:2012rh, Hannam:2013oca, Khan:2018fmp, Khan:2015jqa, Khan:2019kot}; instead in the plunge-merger-ringdown regimes, the deviation coefficients are applied to phenomenological quantities extracted from complete numerical solutions of the Einstein equations, within the Phenom family. The expansion of Eqs. \ref{eq:generic_expansion} was instead used both in the case of the parametrized-EOB~\cite{Brito:2018rfr} model, including in the analysis the whole signal, or in the \texttt{pyRing} approach~\cite{CARULLO-DELPOZZO-VEITCH, ISI-GIESLER-FARR-SCHEEL-TEUKOLSKY}, which in order to avoid contamination from earlier stages of the coalescence, formulates the problem completely in the time domain and restricts the analysis to the post-merger-only section.\\
The precise set of theories and the parameter space explored within each alternative theory by such a parametrization depends on the specific choice of prior bounds used, and is difficult to compute due to the lack of accurate or complete inspiral-plunge-merger-ringdown predictions from alternative theories. As already mentioned, one of the disadvantages of this latter approach is that the deviations parameters can take different values for different sources and it is thus not possible to directly combine the obtained constraints. Such a difficulty has been overcome using a \textit{hierarchical approach}~\cite{Hierarchical_I, Hierarchical_II}, which places constraints on the underlying \textit{observed population distribution} of such deviations (a Dirac distribution centered around the null value, if GR is correct and the measurement has infinite precision).
This approach is the optimal -- following directly from Bayes theorem -- procedure of "stacking" information from multiple sources in the absence of additional hypotheses on the underlying alternative theory of gravity. These tests also allow the intrinsic (masses and spins) parameters of the binary to deviate from the values which can be extracted by assuming that GR is the correct description of the signal. This feature is required since without a perturbative formulation the intrinsic parameters entering the waveform are $(\bar{M}, \bar{\chi})$, not $(M, \chi)$.
Once again, this has the effect of allowing for more freedom in the reconstructed signal, but it adds a large amount of correlation among the intrinsic binary parameters and the deviation parameters. As ubiquitous in statistics, the gain in generality that these tests have in both allowing non-perturbative deviations in the waveform parameters and in considering deviations in the mass and spin parameters degrades the sensitivity of the tests.\\
The ParSpec formalism instead models possible deviations from GR in a perturbative spirit. Deviations in both binary parameters and coefficients determined by the underlying dynamics are considered within expansions with respect to the GR value. An implication of this approach is that now the expansion can be formulated in terms of the GR binary parameters and consequently these parameters can be restricted within the support of the posterior probability distribution obtained when assuming GR, strongly reducing the amount of correlation present in the problem. This approach restricts the deviations values considered, but strongly enhances the resolving power of a test, allowing to detect deviations with a much smaller number of sources that would be needed by the more generic tests considered by the LIGO-Virgo collaborations. Furthermore, as already discussed above, the restrictions applied in the parameter space do not affect the capabilities of the framework to \textit{detect} a violation from GR even for scenarios where non-perturbative signal deviations are present.\\

\section{ParSpec: high-spin version}
\label{sec:ParSpec_highspin}

In the original reference where the ParSpec formalism was proposed~\cite{ParSpec}, the authors provided fitting coefficients of the GR QNM complex frequencies valid only up to adimensional spin $a_{max}=0.7$. This value was chosen being the typical value of a BH remnant formed from the quasi-circular coalescence of two mildly spinning, close to equal mass BHs, the typical sources from which ringdown is observed.
However, the low-SNR of current LVC observations implies that, although the posterior distributions on the spin parameter typically peak at values close to $0.7$,   they often have support up to high spins ($a \sim 0.95$)~\cite{O3a_TGR}. Consequently, before proceeding to analyse LVC data, we need to perform an extension of the fits provided in Ref.~\cite{ParSpec} in order to faithfully fit QNM numerical data up to larger values of the spin. As a conservative choice, we choose to extend the fits up to $a_{max}=0.99$.
This extension does not require any change to the general formalism in itself, since the accuracy of the expansion contained in Eq.~(\ref{eq:ParSpec_expansion}) increases with the maximum order of the expansion $N_{max}$ and is extendable to arbitrarily large spins.

\subsection{Fit framework}

Numerical difficulties in high-order expansions however do arise from the need of simultaneously fitting a large number of coefficients, particularly if the expansion is carried directly on the damping time $\tau$, which gives rise to a more complex structure with respect to different parametrization such as the quality factor $Q = \frac{\omega_r}{2\omega_i}$, where $\omega_r$ ($\omega_i$) is the real (imaginary) part of the complex QNM frequency.
These difficulties can nonetheless be circumvented by employing sophisticated fitting algorithms able to capture the full multi-dimensional space structure of the -- often strongly -- correlated coefficients. In this work we make use of the \texttt{CPNest}~\cite{cpnest} Nested-Sampling algorithm, able to reconstruct multi-dimensional, strongly correlated posterior distributions (see~\cite{Sivia} for an introduction to Nested Sampling). The algorithm has been validated with a wide range of high dimensional analytic problems (e.g. Rosenbrock function, 50-dimensional gaussian, eggbox profile), by reproducing LIGO-Virgo measurements of compact binary GW observations and is routinely employed in the production of merger-ringdown results by the LIGO-Virgo collaborations~\cite{O3a_TGR}.
The algorithm has also been used to extract post-Newtonian coefficients from full numerical simulations of the Einstein equation, describing the coalescence of two compact objects. Once a suitable and robust parametrization is provided, analytically unknown coefficients can be extracted from numerical data~\cite{Breschi_thesis}. See also~\cite{Messina:2017yjg, Messina:2019uby, vandeMeent:2020xgc} for alternative approaches on similar cases. Applying the algorithm to our problem requires to formulate the fit in terms of a simple inference problem in Bayesian statistics (for a rigorous formulation of Bayesian probability theory as an extension of classical logic, see Ref.~\cite{Jaynes2003}). If the coefficients to be extracted are collectively represented by $\vec{\theta}$, the probability distribution on the coefficients conditioned on the available numerical data $D$, the \textit{posterior distribution}, is computed through Bayes theorem:

\be
    p(\vec{\theta} | d, \mathcal{H}, I) = \frac{p(\vec{\theta} | \mathcal{H}, I) \cdot p(d | \vec{\theta}, \mathcal{H}, I)}{p(d | \mathcal{H}, I)} \,, 
\ee
where $\mathcal{H}$ represents the specific parametric template describing the data (our hypothesis) and $I$ denotes all the available background information.
The \textit{prior distribution} $p(\vec{\theta} | \mathcal{H}, I)$ encodes all the information available on the coefficients before the inference process (e.g. positive defined, bounded by some physical constraint, etc.). If no a-priori information is available, the prior can be conservatively chosen to be uninformative. The \textit{likelihood} $p(d | \vec{\theta}, \mathcal{H}, I)$ is determined by the error distribution of the available numerical data. The normalisation term $\mathcal{Z} := p(d | \mathcal{H}, I)$, the \textit{evidence}, quantifies the probability that the data $d$ can be explained by the model under consideration. The name \textit{evidence} can be understood by the role of this term when comparing the agreement with the data of two competing models $\mathcal{H}_1, \mathcal{H}_2$:
\be
    \frac{p(\mathcal{H}_1 | d, I)}{p(\mathcal{H}_2 | d, I)} = \frac{p(\mathcal{H}_1 | I) \, p(d | \mathcal{H}_1, I)}{p(\mathcal{H}_2 | I) \, p(d | \mathcal{H}_2, I)} := \frac{p(\mathcal{H}_1 | I)}{p(\mathcal{H}_2 | I)} \cdot \mathcal{B}_{2}^{1} \,, 
\ee
The first term in the above equation represents the ratio of prior probabilities of the models; the second term, the \textit{evidences ratio}, is called \textit{Bayes factor}. When comparing GR predictions against those of a specific modified theory of gravity, we will quote this number.
The advantages of such an approach with respect to standard $\chi^2$ methods mainly consist in: i) providing the full multi-dimensional probability distributions on the parameters, encoding the global correlation structure of the problem and thus going beyond simple mono-dimensional error estimates on each of the parameters; ii) avoid overfitting by using the Bayesian evidence as a tool to select the best number of parameters needed to fit the dataset\footnote{Although the latter point is important in principle, it is only marginally relevant for the problem at hand, since we are interested in an effective low-order polynomial representation of the complex ringdown frequencies. Overfitting issues would be relevant only for very high order expansions saturating the amount of information contained in the numerical data. As shown below, current statistical uncertainties keep the number of needed coefficients small enough that overfitting is not presently a concern.}.\\ 

\subsection{Results}

We use Eq.~(\ref{eq:ParSpec_expansion}) as a template, hence defining our hypothesis $\mathcal{H}$ and apply Bayes Theorem to extract posterior probability distributions on the parameters $\vec{\theta} = \{ \omega^{(j)}, \tau^{(j)} \}_{j=1,...,N_{max}}$. The dataset $d$ employed is the one of Ref.~\cite{Berti_fits, Berti_website}, consisting in $10^4$ entries of Kerr QNM frequencies and damping times for values of the BH spin in the range $[0, 0.9999]$, obtained numerically through the continued fraction method of Leaver~\cite{Leaver:1985ax}. We consider the modes relevant to current or near-future observational analyses, namely $(l,m,n) = \{(2,2,0), (2,2,1), (3,3,0), (2,1,0) \}$. We set uniform priors on all the coefficients with bounds [-100,100].
The likelihood function is determined under the assumption that the error on the numerical data is distributed as a zero-mean gaussian with a conservative standard deviation estimate of $10^{-5}$, well below the precision achievable by current merger-ringdown data analyses.
The order of the expansion $N_{max}$ is dictated by the required precision in representing the data throughout the full spin range. Near-future expected statistical errors on the $\delta\omega$ and $\delta\tau$ parameters are of the order of few percent for the frequency and a few tens of percent for the damping time, given that this parameter is only weakly measured~\cite{Meidam}; thus, as a conservative choice, we keep including terms in the expansion until the residuals go below $1\%$ ($3\%$). The zeroth-order coefficients are fixed from the known Schwarzschild values. The sampler settings used are: 2048 live points, 1024 as a maximum number of Markov-Chain Montecarlo internal steps and a pool-size composed of 256 walkers. The stopping condition is set by requiring an estimated precision of 0.1 on the logarithm of the evidence.\\

Obtained results are collected in Table \ref{tab:coeffs_table}, where we show median and $90\%$ credible level values for the $N_{max}=5 \, (9)$ expansion order, which provides residuals below $1\% (3\%)$ on the $\omega (\tau)$ fitting parameters.
For high expansion orders, $\tau$ fitting parameters tend to hit the chosen prior bounds (we checked that wider bounds do not alleviate the problem), showing large cancellations among the different orders. Such behaviour is tipically observed when polynomial expansions are used to represent highly-structured functions. This is not a concern, because we only need an effective expansion which is accurate enough for our purposes; the fact that polynomial expansions are a poor fit of QNM complex frequencies has long been known and indeed the most accurate parametric representations of these data tipically involve non-rational functions~\cite{Berti_fits, Nagar_fits, London_fits}. A polynomial expansion has the advantage of being easily interpretable when analysing observations, but most importantly to be easily generalisable to arbitrary alternative theories with additional charges. As a future development it would be interesting to explore other parametrizations with respect to the simple polynomial one considered here and in Ref.~\cite{ParSpec}.

\begin{table*}[t]
\caption{Numerical results for the coefficients of the polynomial expansion considered in Eq.~(\ref{eq:ParSpec_expansion}). The $c_{n}$ coefficients for a given mode correspond to $\omega^{(j)}_K, \tau^{(j)}_K$ coefficients in Eq.~(\ref{eq:ParSpec_expansion}), for $K=\{ (2,2,0), (2,2,1) , (2,1,0), (3,3,0) \},  j=0,\cdots,9$. The non-spinning coefficients are obtained from perturbation theory of Schwarzschild BHs.\\}
\begin{ruledtabular}
\resizebox{0.8\textwidth}{!}{\begin{tabular}{c|cccc}
 & & \qquad \qquad \qquad \qquad $\omega$ & & \\
\hline \hline
(l,m,n) & $(2,2,0)$ & $(2,2,1)$ & $(2,1,0)$ & $(3,3,0)$ \\
\hline \hline
c$_0$ & ${0.373672}$                    & ${0.346711}$                    & ${0.373672}$                      & ${0.599443}$                    \\
c$_1$ & ${0.2438}^{+0.0001}_{-0.0001}$  & ${0.2605}^{+0.0001}_{-0.0001}$  & ${0.08275}^{+0.00009}_{-0.00009}$ & ${0.3640}^{+0.0001}_{-0.0001}$  \\
c$_2$ & ${-1.2722}^{+0.0008}_{-0.0009}$ & ${-1.2529}^{+0.0009}_{-0.0008}$ & ${-0.1860}^{+0.0008}_{-0.0007}$   & ${-1.7372}^{+0.0007}_{-0.0009}$ \\
c$_3$ & ${4.925}^{+0.002}_{-0.002}$     & ${4.875}^{+0.002}_{-0.002}$     & ${0.880}^{+0.002}_{-0.002}$       & ${6.762}^{+0.002}_{-0.002}$     \\
c$_4$ &$ {-7.031}^{+0.003}_{-0.003}$    & ${-6.968}^{+0.003}_{-0.003}$    & ${-1.265}^{+0.002}_{-0.003}$      & ${-9.649}^{+0.002}_{-0.003}$    \\
c$_5$ & ${3.621}^{+0.001}_{-0.001}$     & ${3.598}^{+0.001}_{-0.001}$     & ${0.692}^{+0.001}_{-0.001}$       & ${4.969}^{+0.001}_{-0.001}$     \vspace{0.05cm}\\
\hline
 &      & \qquad \qquad \qquad \qquad $\tau$ & & \\
\hline \hline
(l,m,n) & $(2,2,0)$ & $(2,2,1)$ & $(2,1,0)$ & $(3,3,0)$ \\
\hline \hline
c$_0$ & ${11.240715}$                    & ${3.650769}$                     & ${11.240715}$                    & ${10.787132}$                     \\
c$_1$ & ${2.3569}^{+0.0001}_{-0.0001}$   &  ${-2.1606}^{+0.0001}_{-0.0001}$ & ${-2.3600}^{+0.0001}_{-0.0001}$  & ${2.35574}^{+0.00006}_{-0.00007}$ \\
c$_2$ & ${-5.0014}^{+0.0004}_{-0.0004}$  & ${27.194}^{+0.001}_{-0.001}$     & ${30.147}^{+0.001}_{-0.001}$     & ${-4.9954}^{+0.0003}_{-0.0003}$   \\
c$_3$ & ${-37.2271}^{+0.0007}_{-0.0007}$ & ${-86.047}^{+0.004}_{-0.004}$    & ${-97.250}^{+0.005}_{-0.005}$    & ${-36.8058}^{+0.0006}_{-0.0004}$  \\
c$_4$ & ${100.00}^{+0.00}_{-0.01}$       & ${56.269}^{+0.005}_{-0.005}$     & ${75.045}^{+0.006}_{-0.006}$     & ${100.00}^{+0.00}_{-0.01}$        \\
c$_5$ & ${31.2037}^{+0.0006}_{-0.0005}$  & ${100.00}^{+0.00}_{-0.01}$       & ${100.00}^{+0.00}_{-0.01}$       & ${31.1889}^{+0.0005}_{-0.0004}$   \\
c$_6$ & ${-100.00}^{+0.01}_{-0.00}$      & ${-46.110}^{+0.003}_{-0.003}$    & ${-64.299}^{+0.004}_{-0.003}$    & ${-100.00}^{+0.01}_{-0.00}$       \\
c$_7$ & ${-100.00}^{+0.01}_{-0.00}$      & ${-100.00}^{+0.01}_{-0.00}$      &  ${-100.00}^{+0.01}_{-0.00}$     & ${-100.00}^{+0.01}_{-0.00}$       \\
c$_8$ & ${26.8440}^{+0.0002}_{-0.0002}$  & ${-42.553}^{+0.001}_{-0.001}$    & ${-31.076}^{+0.001}_{-0.001}$    & ${26.8726}^{+0.0002}_{-0.0003}$   \\
c$_9$ & ${100.00}^{+0.00}_{-0.01}$       & ${100.00}^{+0.00}_{-0.01}$       & ${100.00}^{+0.00}_{-0.01}$       & ${100.00}^{+0.00}_{-0.01}$        \\
\end{tabular}}
\end{ruledtabular}
\label{tab:coeffs_table}
\end{table*}

\section{Observational constraints}
\label{sec:Observations}

\subsection{Comparison to future detector observations and current expected precision}

Before applying the high-spin version of ParSpec  to observational data, it is instructive to estimate the level of precision we expect to achieve on deviation coefficients, by extrapolating the estimates obtain in Ref.~\cite{ParSpec}. In the latter study, the ParSpec formalism was applied to a set of simulated signals expected to be detected by the Einstein Telescope (ET) interferometer and the LISA mission~\cite{ET, LISA}. The width of the obtained probability distributions on $\delta\omega_0$ were of $\sigma_{90} \sim 3 \times 10^{-3}$, at $90\%$ credible levels, when allowing only a single deviation parameter and combining 10 sources. \\
In this work, we instead apply ParSpec to the only currently available ringdown observations, the post-merger phase of binary BHs coalescences observed by the LIGO and Virgo interferometers. The signal-to-noise ratio ($\mathrm{SNR}$) of these sources lies in the range $5-15$, much smaller compared to expected SNR values from ET/LISA observations, $\mathcal{O}(100-1000)$ as estimated by Ref.~\cite{ParSpec}. Consequently, given that uncertainties scale as $\sim 1/\mathrm{SNR}$, we expect to be able to place constraints of the order of $3 \times 10^{-2}$ on $\delta\omega_0$, under the same hypothesis on the number of deviation parameters and number of combined sources. In a realistic setting, as the one being considered in this work, discrepancies from this simple estimate arise from: i) poor accuracy of the the aforementioned SNR scaling in the low-SNR limit; ii) imperfect wide-sense stationarity or gaussianity of the stochastic process governing the interferometer output, and iii) the marginalisation of additional parameters other than parametric deviations. The last point may also introduce, through correlations, deviations in the gaussian approximation (e.g. wider tails) of the posterior probability distributions made in Ref.~\cite{ParSpec}. All these features will contribute to widen the posterior probability distribution and imply that the estimate presented above can only represent an optimistic lower bound on the precision we can obtain. When comparing constraints drawn from observations of stellar-mass BHs (GW ground-based detections) against those obtained from supermassive BHs (GW space-based detections or BH shadows observations), it should be recalled that these different mass scales actually probe different curvature regimes. Although ringdown signals genererated by mergers of BBH in LISA will be loud, the corresponding space-time curvature (quantified by the Kretschmann invariant) associated to the emitters, massive BBHs, will be smaller compared to the one associated with stellar-mass BH mergers observed by current and future ground-based detectors. Given that GR deviations are expected to be prominent in the high-curvature regime, the high-frequency (larger than kHz) regime~\cite{NEMO} will probe the most interesting regions of the curvature parameter space. Observations of post-merger signals from low-mass compact objects will also unveil the formation mechanisms of remnants lying on the border of the BH low-mass gap, extending spectroscopic measurements to ECOs that may avoid gravitational collapse only in a limited mass range or even to primordial BHs mergers~\cite{Cardoso:2019rvt}. In summary, constraints placed by space and ground-based detectors should be regarded as complementary in the curvature parameter space (for a comprehensive review of the possible constraints on alternative theories of gravity when considering future ground and space-based detectors, see Ref.~\cite{Perkins:2020tra}). Planned ground-based detectors such as the Neutron Star Extreme Matter Observatory (NEMO)~\cite{NEMO}, targeting GW observations in the kHz-band, could soon open an observational window onto this exciting frequency band.

\subsection{Analysis infrastructure: \texttt{pyRing}}

To constrain deviation coefficients from Eq.~(\ref{eq:ParSpec_expansion}), we use \texttt{pyRing}~\cite{CARULLO-DELPOZZO-VEITCH, ISI-GIESLER-FARR-SCHEEL-TEUKOLSKY}, a publicly available ringdown-tailored parameter estimation \texttt{python}~\cite{python} package, employing a Bayesian approach and a fully time-domain formulation of the inference problem, both in the likelihood and in the waveform. The package is based on the \texttt{CPNest}~\cite{cpnest} sampler and relies on the \texttt{LALInference} library~\cite{lalinference} to compute projections onto detectors. The package internally implements most of the state-of-the-art merger-ringdown-only analytical waveform templates available in the literature~\cite{Kamaretsos2012, MMRDNS, MMRDNP, Damour_Nagar_ring} and accesses other standard inspiral-plunge-merger-ringdown waveforms thorugh the \texttt{LALSimulation} library~\cite{lalsuite}. Both ringdown-only template injections and complete inspiral-plunge-merger-ringdown injections can be performed; the latter are available both for analytical templates and for numerical simulations accessed through the \texttt{LALSimulation} library and the LIGO-Virgo Numerical Relativity Injection Infrastructure~\cite{Galley:2016mvy, Schmidt:2017btt}. To optimise computations, \texttt{pyRing} internal waveforms and all likelihood calculations are implemented in \texttt{cython}~\cite{cython}. Computations involving the Auto-Covariance matrix are drastically optimised using some of the state-of-the-art algorithms introduced in Ref.~\cite{Geophys}, exploiting recursive relations obtained from the Toeplitz structure of the Auto-Covariance matrix. QNM frequencies can be pre-computed using \texttt{SciPy}~\cite{scipy} cubic-spline interpolation of publicly available numerical data from Ref.~\cite{Berti_fits}, publicly avaialable at~\cite{Berti_website}.
Remnant BH mass and spin are computed using the fits presented in Ref.~\cite{JIMENEZ-FORTEZA-KEITEL-HUSA-HANNAM-KHAN-PUERRER, surfinbh1, surfinbh2} and accessed through the \texttt{LALSimulation} library or the correspective \texttt{python} packages. The \texttt{pyRing} package was recently used to produce the first catalog of ringdown-only remnant properties measurements and constraints on QNM complex frequencies parametric deviations, using data from the first three observing runs of the LIGO-Virgo interferometers, see Tables VIII-IX of Ref.~\cite{O3a_TGR}. We adopt the same selection criteria of Ref.~\cite{O3a_TGR} to select the list of events on which we apply the ParSpec formalism, namely all the signals in the GWTC-2 catalog with a False Alarm Rate smaller than $10^{-3}$ per year as reported by any of the search pipelines employed and which additionally present evidence for a ringdown signal as quantified by the Bayes Factors against the hypothesis that the data contained only gaussian noise. Since low-mass events merge at high frequencies where the sensitivity of the detectors is lower and thus the ringdown is not detectable, this subset of events essentially corresponds to the massive branch of loud signals reported in the GWTC-2 catalog.
When considering $\delta\tau$ deviations, we exclude from the analyses the events: GW170814, GW190521, GW190828$\_063405$. The inference on the damping time, a parameter currently only weakly measured, of these events has in fact been shown to suffer from systematic errors due to behaviour of the stochastic process governing LIGO-Virgo noise~\cite{LIGOScientific:2019fpa, O3a_TGR}. 
This choice sets the number of events here considered to 17 (14 when considering $\delta\tau$ parameters) out of a total of 50 BBH events detected by the LIGO-Virgo collaborations, minimising the effect of noise mimicking a GW event and contaminating tests of GR. The list of events included is summarised in Table~\ref{tab:GWTC2_events}. Also, we employ the same strain data, data conditioning methods and auto-correlation-function parameters of Ref.~\cite{O3a_TGR}. All these settings and the corresponding analyses results are publicly available from the accompanying data release~\cite{O3a_TGR_data_release}.
The waveform template used is $\mathrm{Kerr_{221}}$, a superposition of the longest-lived mode and its first overtone ($n=0,1$) with free complex amplitudes, for details see Refs.~\cite{Giesler:2019uxc, ISI-GIESLER-FARR-SCHEEL-TEUKOLSKY, O3a_TGR}. The analysis segment starts at the peak of $h_+^2 + h_{\times}^2$ in each detector. Complex frequencies are expressed using the fits of Eq.~(\ref{eq:ParSpec_expansion}), extended to large spin values as discussed in Sec.~\ref{sec:ParSpec_highspin}. 
Consistently with the ParSpec perturbative approach, we set uniform bounds on all deviation parameters in $[-0.5,0.5]$ range, on the $\ell$ scale in the $[0,75]\, \mathrm{km}$ interval and marginalise the remnant mass and spin within the 90\% CLs of their GR values. To break the strong degeneracy with the distance (which is poorly measured for merger-ringdown-only signals) in the dimensionful $\alpha$ case, we also restrict it to its 90\% GR CLs interval and apply a uniform prior within this interval. All the other priors and analysis parameters are chosen to be identical to those of Ref.~\cite{O3a_TGR}. 
In a realistic scenario, it must be recalled that non-zero values of deviation parameters can be caused not only by violations of GR, but from violations in any of the underlying analysis' hypotheses. One of such examples would be an incorrect description of the detector noise or missing physics predicted by GR, but not taken into account into our waveform model (e.g. eccentricity, extreme precession, environmental effects). As verified in Ref.~\cite{O3a_TGR}, all of these assumptions hold for current LIGO-Virgo observations.\\

\begin{table}
\caption{Summary of the events included in the analysis of frequency and damping time deviations. See the main text for a detailed description of the events selection criteria.}
\vspace{0.2cm}
\begin{tabular}{@{}lcc}
\hline\hline
\multicolumn{3}{c}{GWTC-2 ringdown events} \\
\hline
\hline
Event & $\delta\omega$ & $\delta\tau$\\
\hline
GW150914           & \cmark & \cmark \\
GW170104           & \cmark & \cmark \\
GW170814           & \cmark & \xmark \\
GW170823           & \cmark & \cmark \\
GW190408\_181802   & \cmark & \cmark \\
GW190512\_180714   & \cmark & \cmark \\
GW190513\_205428   & \cmark & \cmark \\
GW190519\_153544   & \cmark & \cmark \\
GW190521           & \cmark & \xmark \\
GW190521\_074359   & \cmark & \cmark \\
GW190602\_175927   & \cmark & \cmark \\
GW190706\_222641   & \cmark & \cmark \\
GW190708\_232457   & \cmark & \cmark \\
GW190727\_060333   & \cmark & \cmark \\
GW190828\_063405   & \cmark & \xmark \\
GW190910\_112807   & \cmark & \cmark \\
GW190915\_235702   & \cmark & \cmark \\

\hline\hline
\end{tabular}
\label{tab:GWTC2_events}
\end{table}

\subsection{Combined probability distribution functions: boundary artifacts}

The coupling parameter $\ell$ is positively-defined. This implies that, in the case of a weak measurement where there is support for values of $\ell=0$, its posterior distribution will hit the lower bound of the prior. 
Consequently, when constructing an estimate of the underlying (continuous) probability distribution function from the discrete samples available, it is necessary to take special care in dealing with artifacts at the borders, which could bias the measurement. 
A continuous estimate of each event probability distribution function is needed when producing a combined estimate of the coupling parameter using multiple events, the latter requiring the multiplication of single-event likelihoods (given the statistical independence of the different GW events). Since flat priors are applied on all the parameters of interest, we can employ posterior probability distributions in place of the likelihood functions, renormalising appropriately.
A first step in alleviating border artifacts is to use a logarithmic transformation, for example:

\be
    y(x) = log \left( \frac{x-x_m}{x_M-x} \right) \,, 
\ee

where $x_m, x_M$ represent the lower and upper border of the support considered.
The transformation stretches the distribution boundaries to infinity. 
The resulting transformed distribution then presents a continuous behaviour at the boundaries and producing an estimate free from border artifacts is simpler. The final estimate of the original distribution is then obtained by multiplying the estimate of the transformed distribution with the transformation jacobian $| \frac{dy}{dx} |$.
To compute the continuous estimate of the transformed distribution, we attempted using both a standard Gaussian Kernel Density Estimation (GKDE) method~\cite{scipy} and a Dirichlet Process Gaussian-mixture (DPGM) model, a fully Bayesian non-parametric method. For a pedagogical introduction, as well as an open-source implementation of the DPGM model, see Ref.~\cite{DELPOZZO2018}.
The DPGM model proved superior in alleviating border artifacts, as directly verified on single-event measurements where it provided in all cases a faithful representation of the corresponding samples, and was thus chosen to produce the combined measurements presented in the remainder of the work.

\subsection{Results}

\begin{figure*}[!tb]
\includegraphics[width=0.48\textwidth]{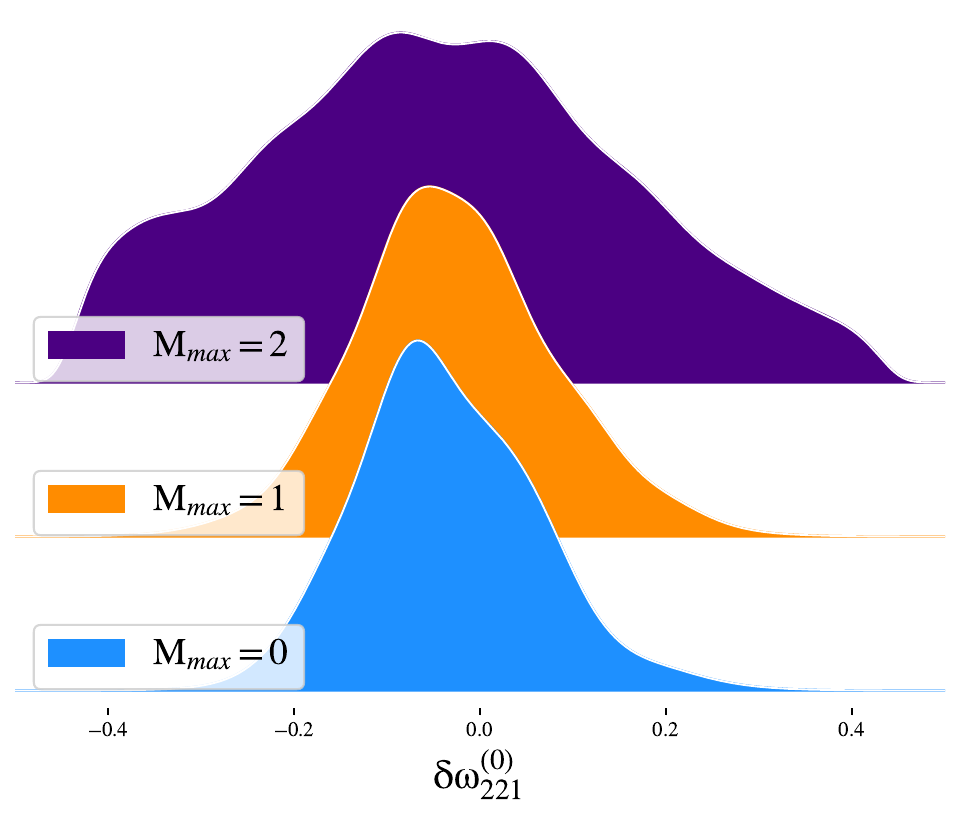}
\includegraphics[width=0.48\textwidth]{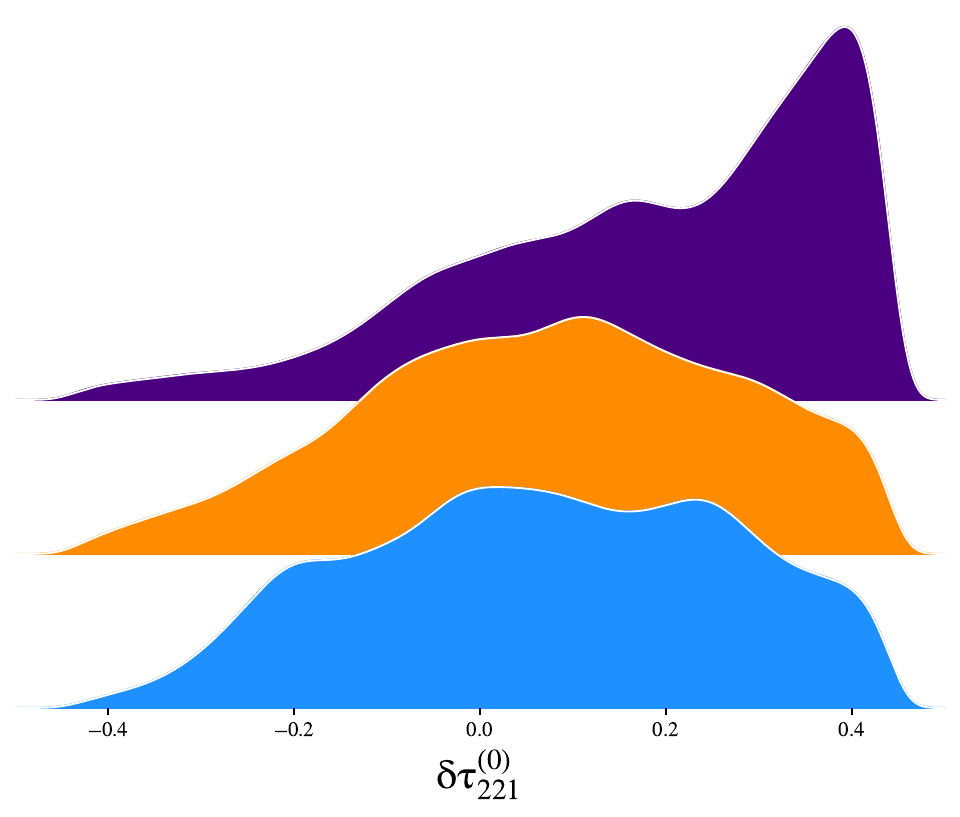}
\includegraphics[width=0.48\textwidth]{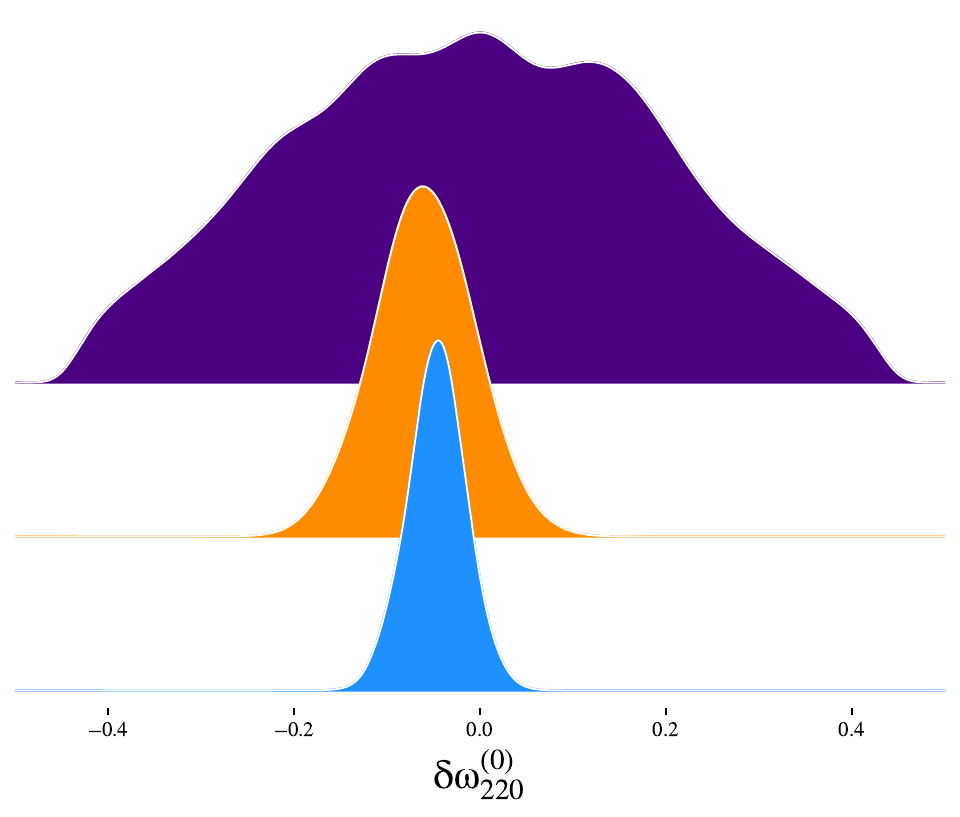}
\includegraphics[width=0.48\textwidth]{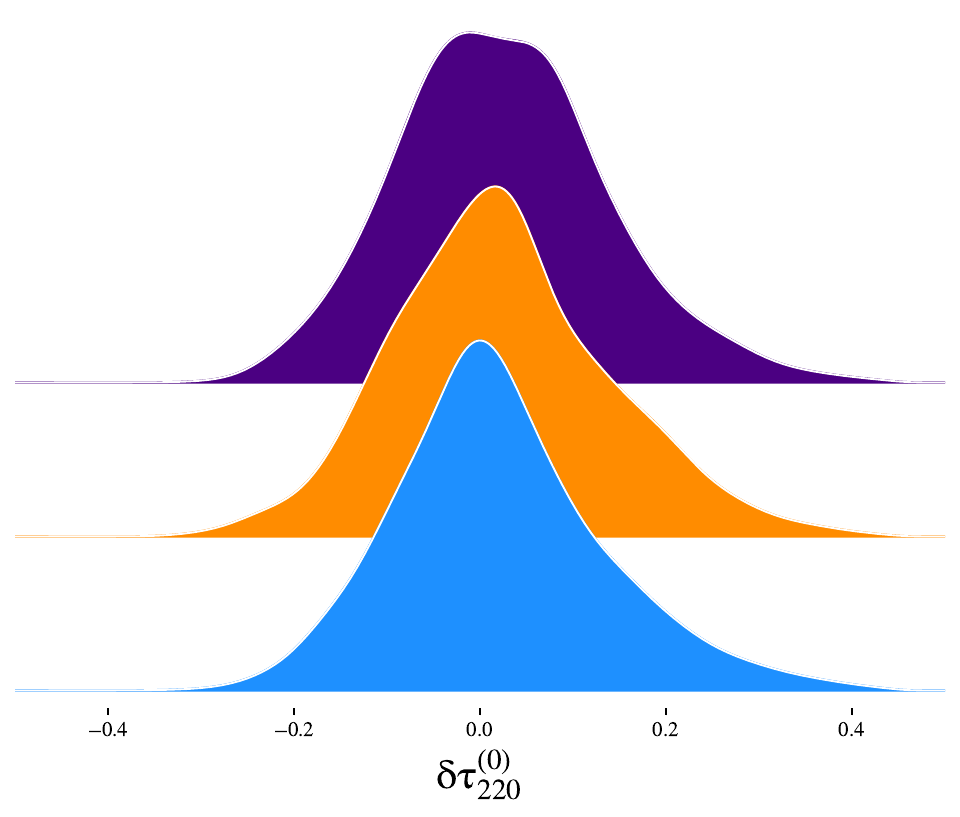}
\caption{Posterior probability distributions on the deviations from the lowest-order (non-spinning) expansion parameters ($\delta\omega, \delta\tau$)  of the $(l=m=2, n=0,1)$ ringdown modes, for the $p=0$ case, when combining all available LIGO-Virgo ringdown observations.}
\label{fig:plot_0}
\end{figure*}

\begin{figure*}[!tb]
\includegraphics[width=0.48\textwidth]{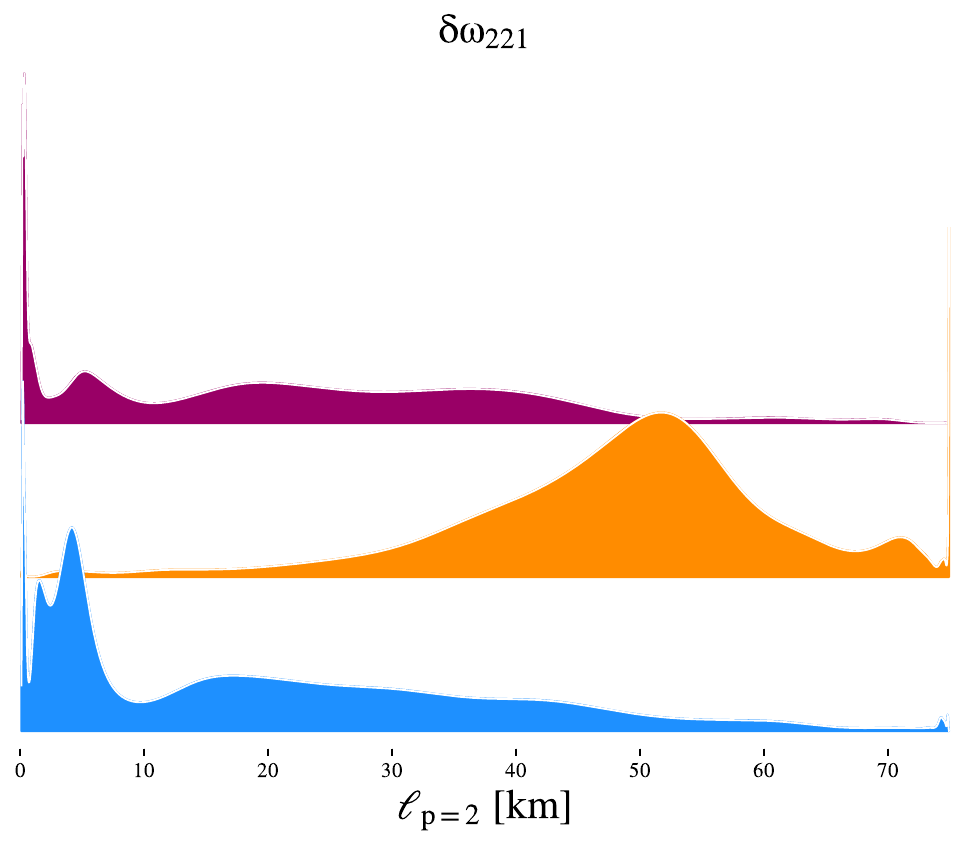}
\includegraphics[width=0.48\textwidth]{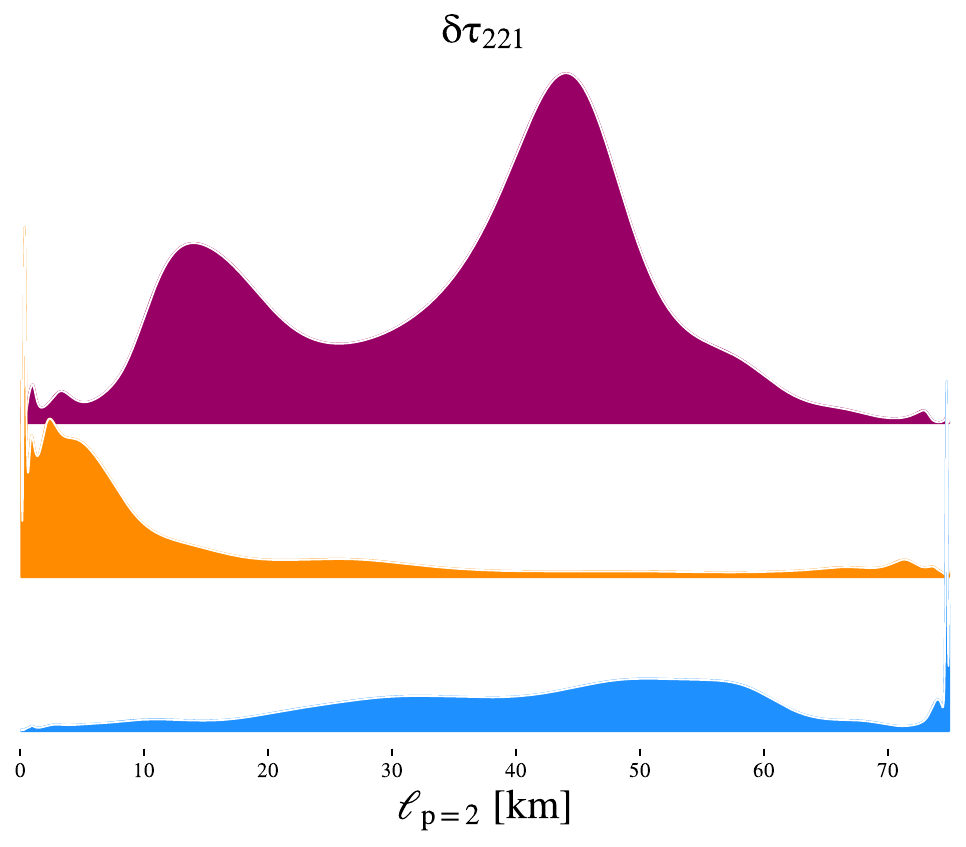}\\
\vspace{0.3cm}
\includegraphics[width=0.48\textwidth]{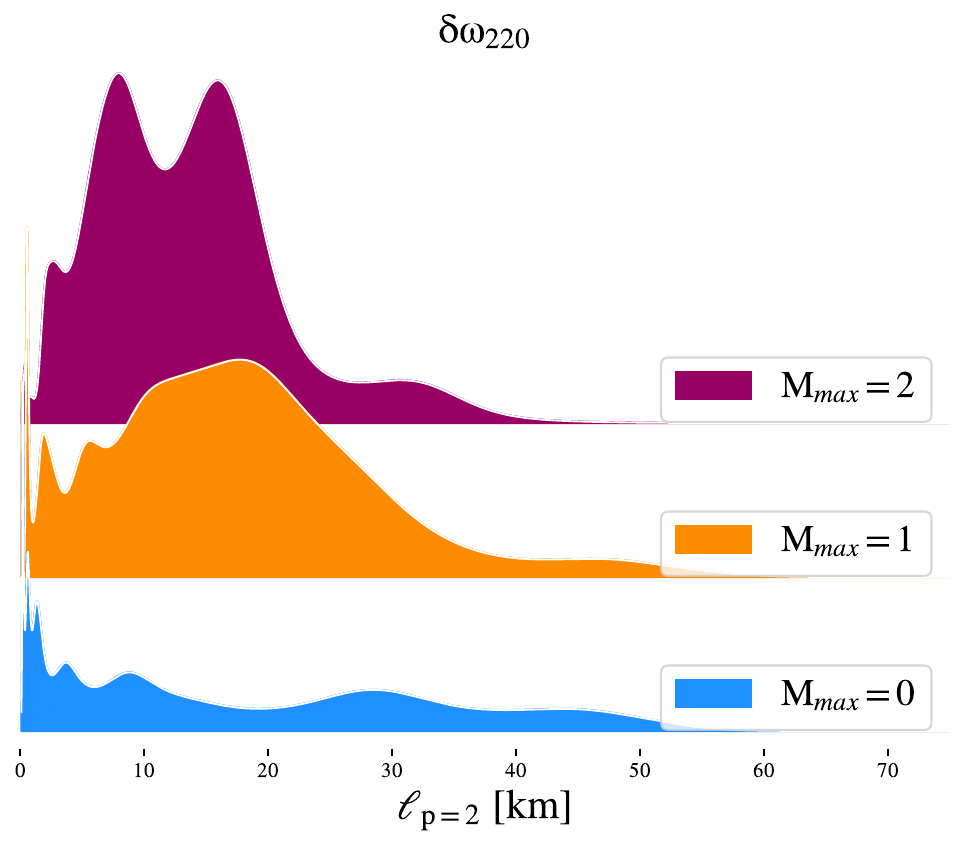}
\includegraphics[width=0.48\textwidth]{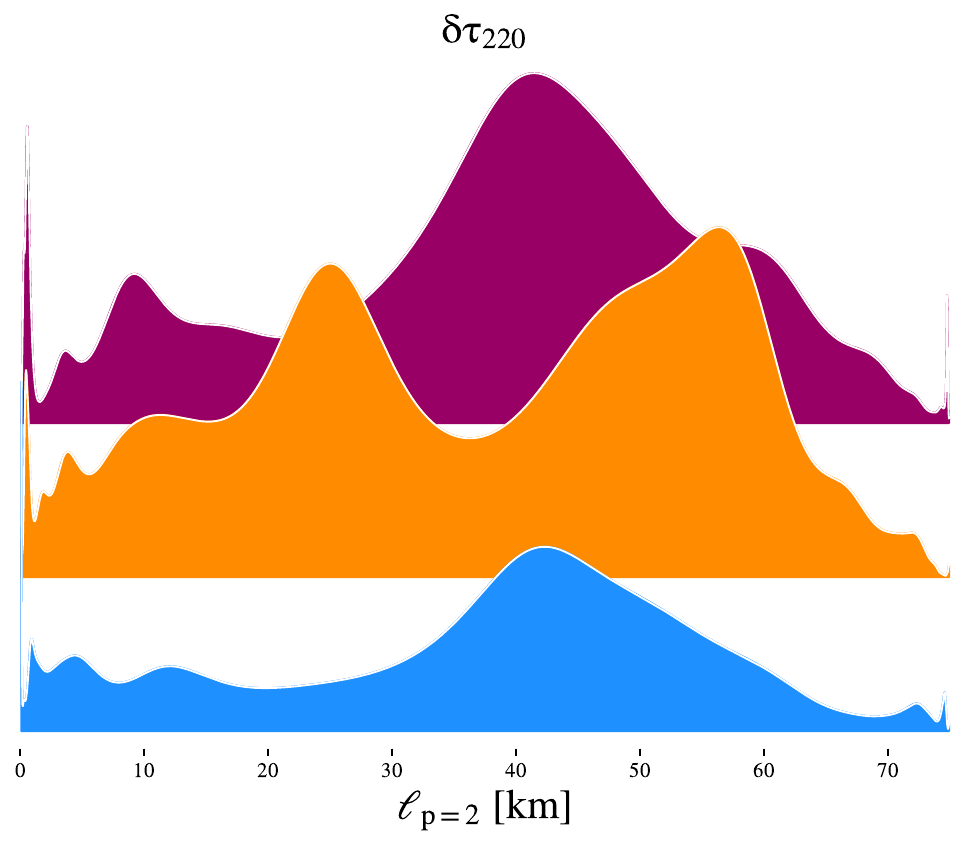}
\caption{Posterior probability distributions on the distance scale $\ell$ below which new physics effects can modify the BH emission process for the $p=2$ case, when combining all available LIGO-Virgo ringdown observations.}
\label{fig:plot_ell_2}
\end{figure*}

\begin{figure*}[!tb]
\includegraphics[width=0.48\textwidth]{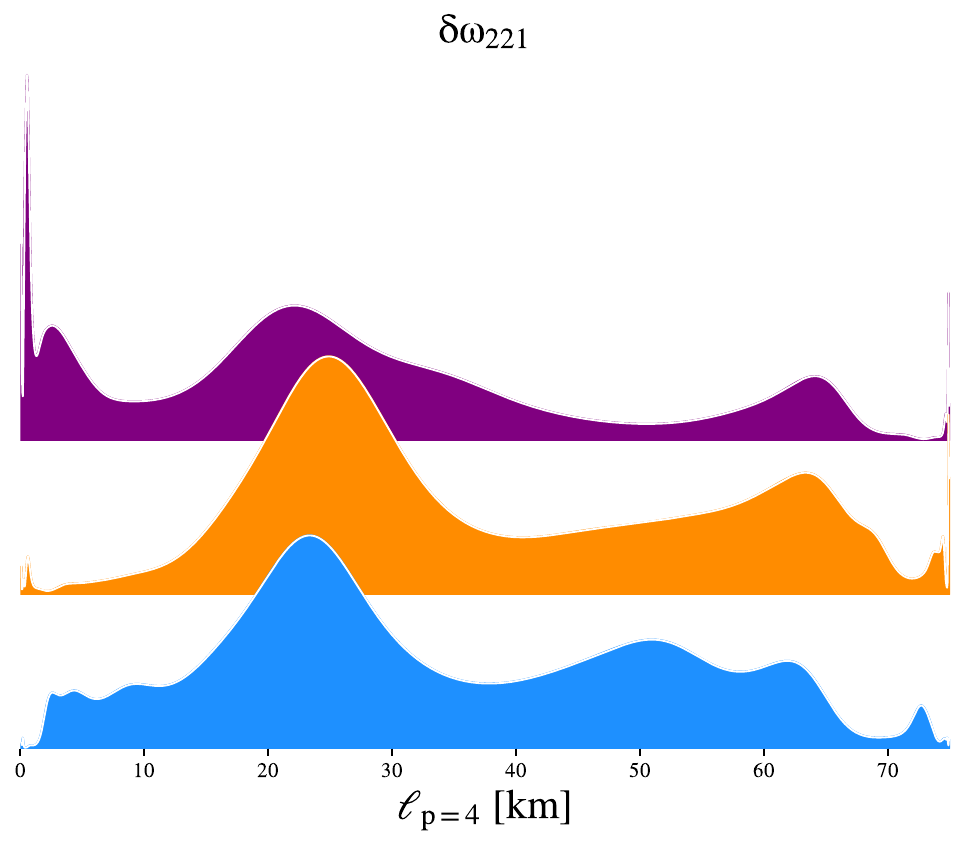}
\includegraphics[width=0.48\textwidth]{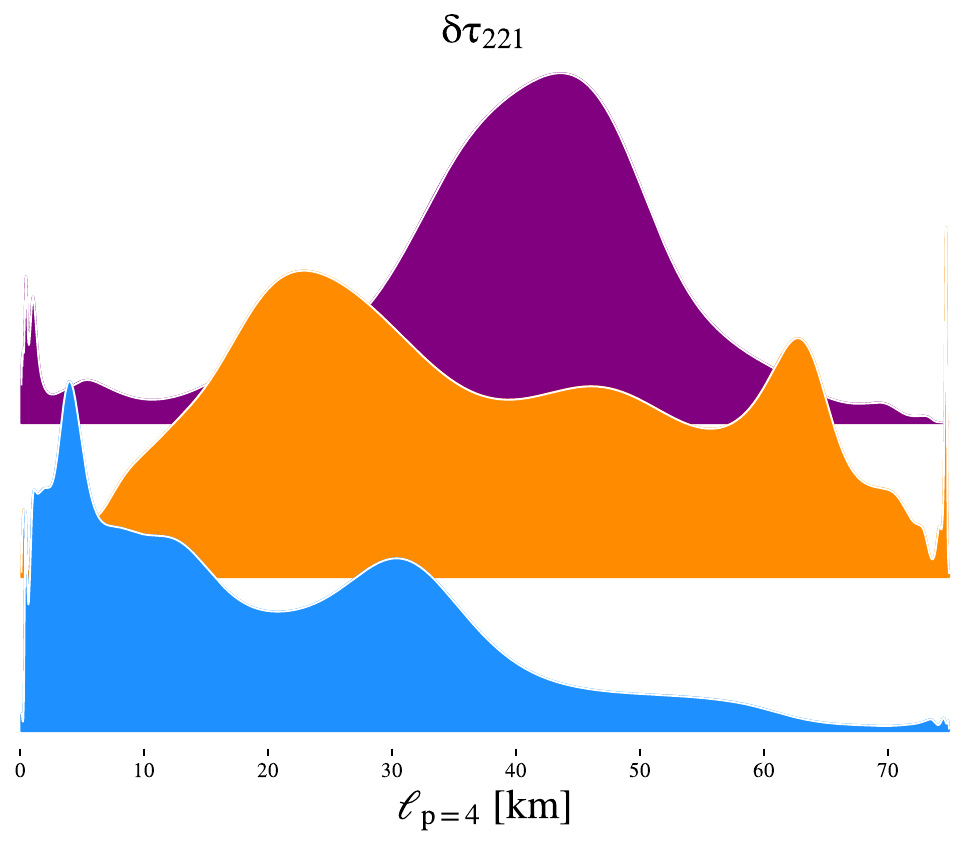}\\
\vspace{0.3cm}
\includegraphics[width=0.48\textwidth]{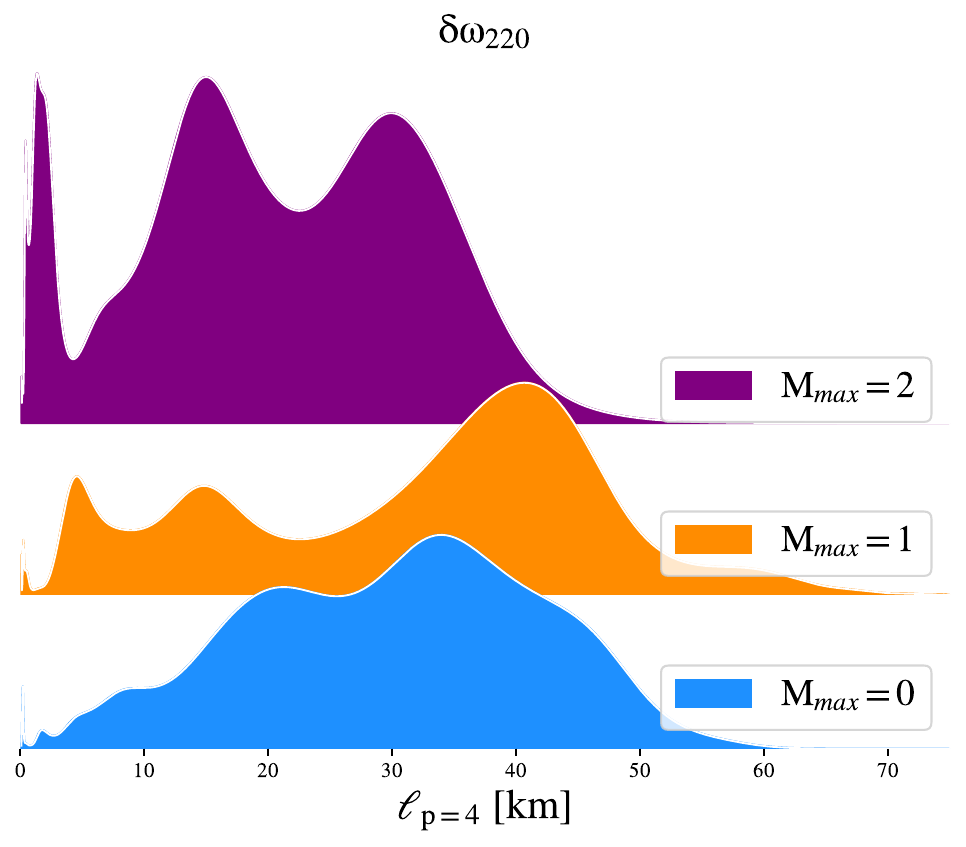}
\includegraphics[width=0.48\textwidth]{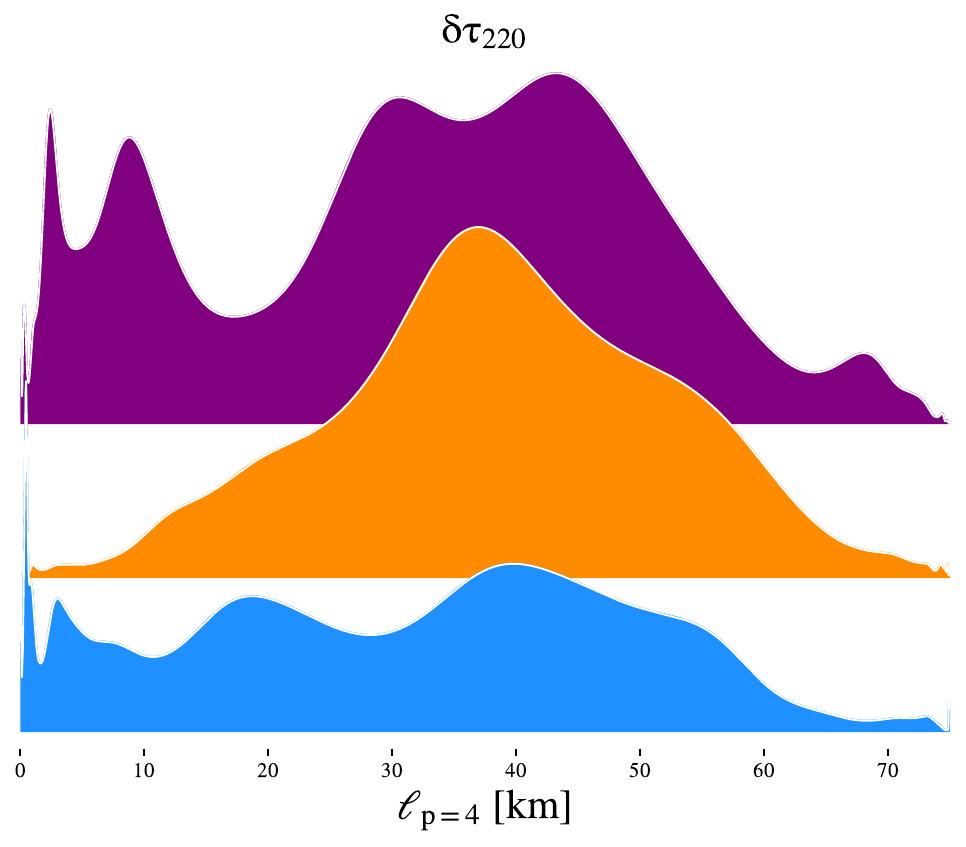}
\caption{Posterior probability distributions on the distance scale $\ell$ below which new physics effects can modify the BH emission process for the $p=4$ case, when combining all available LIGO-Virgo ringdown observations.}
\label{fig:plot_ell_4}
\end{figure*}

\begin{figure*}[!tb]
\includegraphics[width=0.48\textwidth]{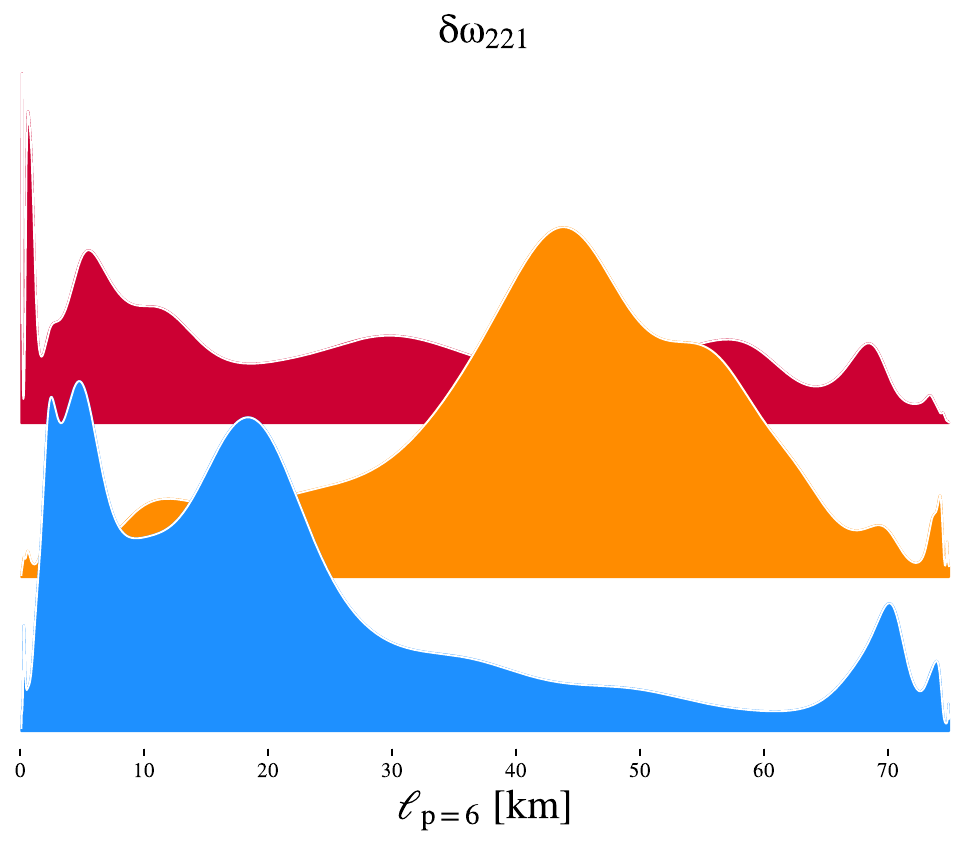}
\includegraphics[width=0.48\textwidth]{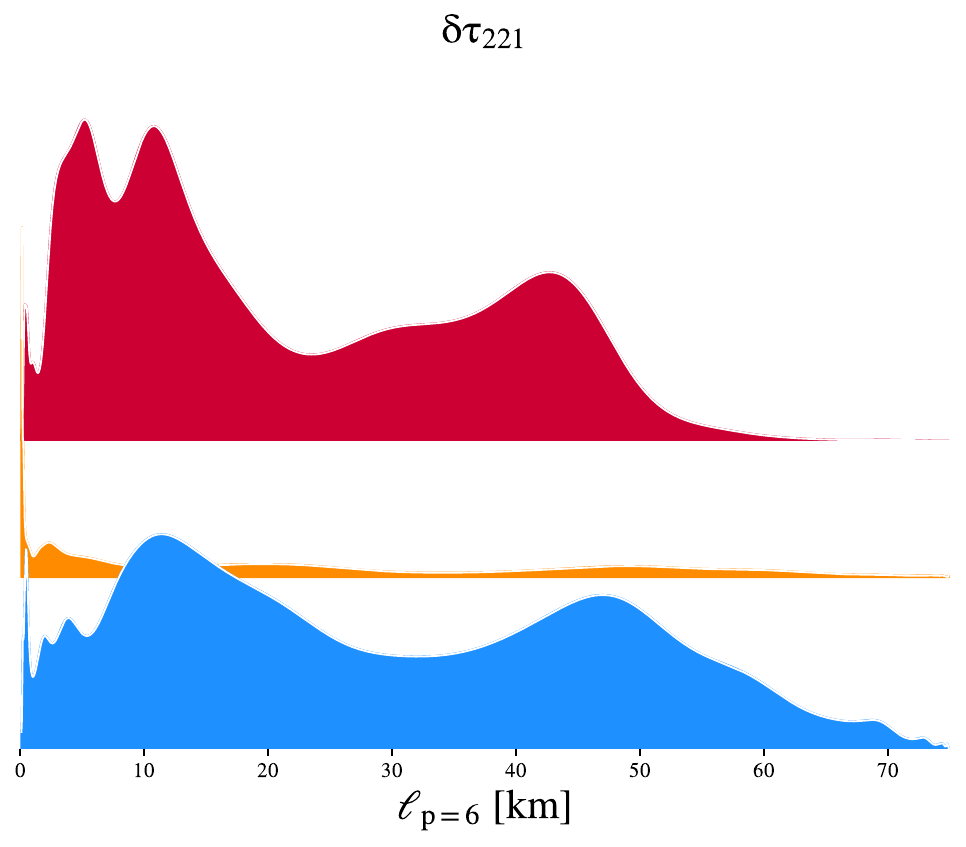}\\
\vspace{0.3cm}
\includegraphics[width=0.48\textwidth]{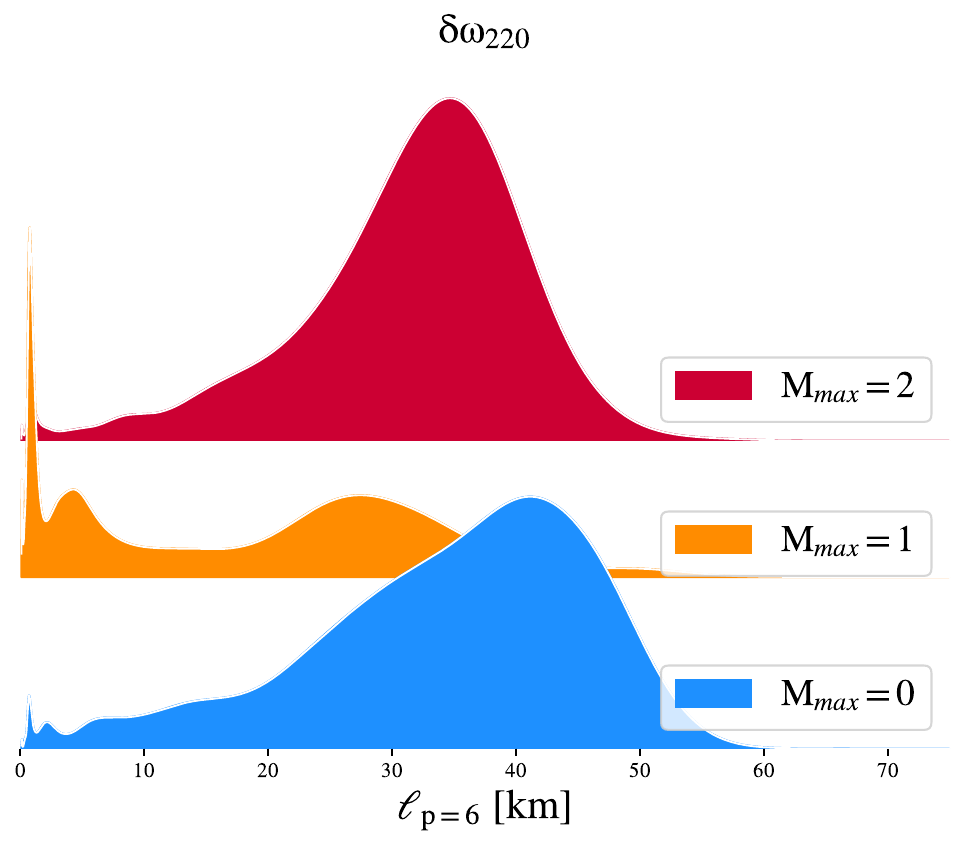}
\includegraphics[width=0.48\textwidth]{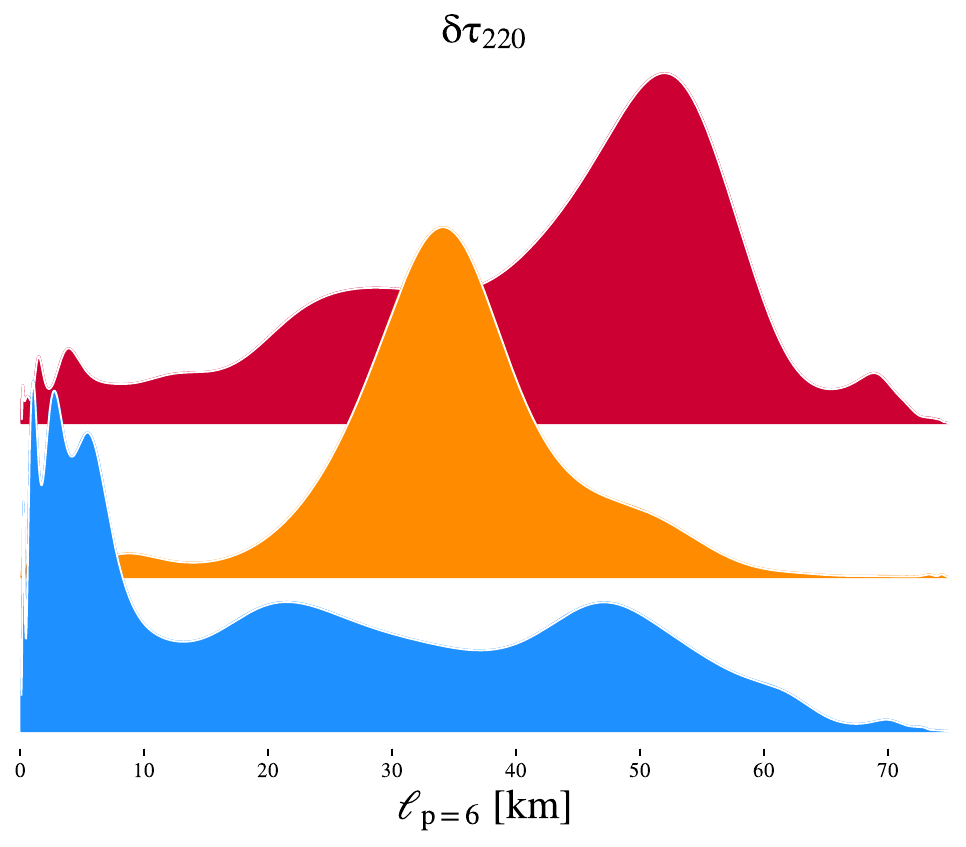}
\caption{Posterior probability distributions on the distance scale $\ell$ below which new physics effects can modify the BH emission process for the $p=6$ case, when combining all available LIGO-Virgo ringdown observations.}
\label{fig:plot_ell_6}
\end{figure*}

We now present constraints on the deviation coefficients and coupling parameter appearing in Eq.~(\ref{eq:ParSpec_expansion}). 
As GR model, we always retain the full GR prediction, keeping the complete expansion up to $N_{max}=5$, for frequency parameters ($N_{max}=9$, for damping time parameters), as determined in Sec.~\ref{sec:ParSpec_highspin}. 
Deviations from GR predictions are instead considered up to a given order $\mathrm{M}_{max}$, while all higher-order deviation parameters are fixed to zero. For each of the charge exponent values $p=0,2,4,6$, we consider $\mathrm{M}_{max}=0,1,2$ . 
Higher orders are limited by computational requirements and the number of available detections\footnote{The approximation induced by the truncation of the expansion at second order has been explicitly validated in Ref.~\cite{Carson:2020cqb} in the context of Einstein-scalar-Gauss-Bonnet gravity, where it was shown that including terms up to the fourth order induced differences much smaller than the current statistical uncertainty on the parameters.}.
In turn we consider deviations on the following set of parameters:$\, \, \{ \omega_{220}, \omega_{221}, \tau_{220},\tau_{221} \}$. Due to computational constraints, we vary coefficients relative to only one of these parameters at a time. When considering $\mathrm{M}_{max} > 0$ we always constraint all lower-order deviation coefficients on the given parameter. As an example, when considering deviations on $\omega_{220}$ for $p=2$ and $\mathrm{M}_{max}=2$, we simultaneously sample on the coefficients: $\{ \ell, \, \delta\omega^{(0)}_{220}, \delta\omega^{(1)}_{220}, \delta\omega^{(2)}_{220} \}$. Results on the deviations parameters of interest are reported in Table~\ref{tab:bounds_table}, for different values of $\{p, \mathrm{M}_{max}\}$, obtained by combining constraints from all the events in the merger-ringdown catalog. 
No statistically significant deviation from the predictions of GR BH dynamics is found.
As expected, tighter constraints can be placed on the $\delta\omega$ coefficients compared to $\delta\tau$ and on the the fundamental ($n=0$) mode compared to the first overtone ($n=1$). The first effect is due to the larger sensitivity of GW detectors to phase deviations with respect to amplitude variations, while the second to the higher SNR carried by the fundamental mode compared to its overtones.
For each mode, the uncertainty on parameters' distributions at different orders is consistent with the magnitude of the correspondent expansion numerical coefficients.
Statistical uncertainties are, as expected, larger than the previously discussed estimates extrapolated from the simplified setting of Ref.~\cite{ParSpec}. 
As also recently discussed in Ref.~\cite{Volkel:2020xlc} in the context of BH shadows observations, considering multiple deviation coefficients at the same time can lead to non-trivial correlations among the coefficients and, depending on the information content of the signal, possibly widen the probability distributions extracted from the analysis. In most cases, if a deviation in the BH dynamics is present, it will affect the expansion at all orders; consequently, to perform a complete analysis all coefficients should be varied at the same time. This, however, needs to be balanced by the limited SNR available, by computational cost of the analysis and by the capabilities of current samplers to handle very high-dimensional problems. Varying coefficients one at a time, as done for example in current parametrized tests by the LVC collaboration, is still a valid method to perform a null test: if no deviations from GR are present, none of the coefficients should peak away from zero. Increasing too much the number of free parameters could mask small deviations by increasing statistical uncertainties.\\
To further highlight this point, in Fig.~(\ref{fig:plot_0}) we show results on the lowest-order (non-spinning) parameters for scalar charges ($p=0$), when considering different $\mathrm{M}_{max}$. The distribution on the same coefficient may widen when higher-order terms are included, as expected due to multi-dimensional correlations. Such degeneracies are difficult to break due to the fact that progenitors with comparable masses and small spins produce remnants with spins $\sim 0.7$. Indeed, events with a detectable merger-ringdown signal reported in the GWTC-2 catalog, present remnant spins in the range [0.6-0.8], as reported in Table VIII of Ref.~\cite{O3a_TGR}. Future observations of events with lower remnant spin (requiring a low-spin, high mass ratio merger or high counter-aligned spins with respect to the orbital angular momentum of the binary) and a detectable merger-ringdown signal will help probing the lower-spin region and further break the aforementioned degeneracies.
To quantify the overall agreement of the observed signals with respect to the predictions of GR, keeping into account the full correlation structure of the signal, we compute Bayes Factors among the violating hypotheses (represented by each of the cases considered in the analysis, as discussed above) and GR. The values are reported in Table~\ref{tab:logBs}.
The Bayes Factors generally decrease with $M_{max}$, penalising the inclusion of more parameters, without support from the data.
They indicate that, when considering scalar ($p=0$) deviations in the fundamental frequency $\delta\omega_{220}$, the data strongly disfavour a violation from GR. Following the reasoning outlined above, as expected stronger constraints can be extracted from the frequencies with respect to the damping times and from the fundamental ($n=0$) modes with respect to the overtone ($n=1$). The remainder of the cases generally show a mild preference for the absence of any GR violation. In all cases no significant deviation from the GR predictions are present within the statistical uncertainty considered.
The combined Bayes Factors are computed under the hypothesis that the same underlying dynamics describes all the observations, with possibly different coupling values for different events, as in the case of a BH possessing an additional $U(1)$ charge (e.g. electric charge). As such, they do not apply to cases of mixed populations of GR BHs and alternative compact objects.
To cover also these cases, we checked the corresponding posteriors and Bayes Factor for each event, none of them showing a statistically significant preference for the violating-GR hypothesis.
Finally, constraints on the coupling parameter $\ell$ for different mass dimensions $p$  and expansion orders $M_{max}$ are presented in Table~\ref{tab:ell_eff_table} and Figs.~\ref{fig:plot_ell_2},\ref{fig:plot_ell_4},\ref{fig:plot_ell_6}.
Since none of the distributions excludes zero, in Table \ref{tab:ell_eff_table} we only report constraints on the charge parameter as $90\%$ upper bounds.\\
Similarly to what has been discussed for the scalar corrections, the strongest bounds come from the fundamental frequency, as expected.
In the informative cases (mainly the ones where deviations in $\delta\omega_{220}$ are considered) the bound correctly shrinks for higher expansion orders, since for higher orders the coupling parameter enters a larger number of terms, increasing its possible impact on the waveform morphology and breaking the correlations with each deviation parameter. 

\subsection{Discussion}

In this section we will put into context our results, comparing them with previous bounds obtained when considering either agnostic parametrizations or specific theories, using both GW and EM observations. When comparing the bounds presented in Tables~\ref{tab:bounds_table},\ref{tab:ell_eff_table} to those obtained through different analyses present in the literature that considered specific alternative theories of gravity, care must be taken into ensuring that we adapt all the results to the same conventions.
In our analysis, $\ell$ is \textit{defined} to be the length scale appearing in the merger-ringdown corrections as specified in Eq.~(\ref{eq:ParSpec_expansion}). Instead, for a given modified theory of gravity, $\ell$ is defined through the coupling constant(s) appearing in the action. These two definitions may thus differ by multiplicative factors, which would also inversely affect our definition of the parametric deviations, since the observable frequency shift is invariant. Also, different authors may sometimes use different definitions of the couplings (again up to a multiplicative constant of order unity). For a review of the different conventions used in the context of Gauss-Bonnet gravity see Ref.~\cite{Witek:2018dmd}.
The units for the quantities obtained in this work are set by the requirement that $\gamma$ corrections enter with constant factors of $O(1)$ and $\delta\omega, \delta\tau$ corrections are comparable with the ones obtained in this work.\\
\textit{\textbf{Scalar coupling:} $\mathbf{p=0}$ --} The stronger constraints imposed by the ParSpec framework translate in a shrinkage of the uncertainties on deviation parameters. For comparison, in the scalar case ($p=0$), when considering a single deviation coefficient, the uncertainty of the lowest-order deviation parameter on the first overtone of the fundamental frequency $\delta\omega_{221}$ is $\sim 2$ times smaller than the result published, under a less stringent parametrization, by the LVC~\cite{O3a_TGR} on a similar parameter. The analogous constraint on the fundamental frequency $\delta\omega_{220}$ is tighter by a factor of 7 than the best known result~\cite{O3a_TGR}, although it should be noted that due to computational requirements, this latter value was obtained when considering only a subset of the full dataset.
Since the uncertainty on the parameters scales roughly as $\mathcal{N}^{-\frac{1}{2}}_{ev}$, $\mathcal{N}_{ev}$ being the number of events analysed, this implies a reduction of a factor $\sim 4$ in the number of events needed to detect a given deviation when using ParSpec, compared to the parametrisation adopted by the LVC.
Additionally, we note that the measurements reported in Table~\ref{tab:bounds_table} show that bounds derived from upcoming LIGO-Virgo-Kagra~\cite{Kagra} observations will soon approach the $\sim \mathcal{O}(1\%)$ limit, as foreseen in Ref.~\cite{Carullo:2018sfu}.\\
\textit{\textbf{Dimensionful coupling:} $\mathbf{p=2}$ --} The bound obtained in this case is $\ell \lesssim 23 \, \mathrm{km}$. The only theory known to us where such a correction is present is a BH possessing a $U(1)$ charge. Given that, in the best motivated theories, this charge is source dependent (as in the case of electrically charged BHs in models of minicharged dark matter~\cite{2016JCAP...05..054C}), the combined constraints presented do not apply to these sets of theories. As such, we lack previous constraints on motivated theories against which we could compare our results.
More in-depth investigations of this specific case will be presented in a future publication.\\
\textit{\textbf{Dimensionful coupling:} $\mathbf{p=4}$ --} In this case we obtain a bound of $\ell \lesssim 35 \, \mathrm{km}$. Two notable examples of quadratic gravity theories with a QNM spectrum featuring $p=4$ corrections, are Einstein-scalar-Gauss-Bonnet (EsGB) and dynamical-Chern-Simons gravity (dCS).\\
Known bounds on the coupling constant of EsGB gravity are summarised in Table II of Ref.~\cite{Perkins:2020tra} and Table I of Ref.~\cite{Okounkova:2020rqw}. In the latter work, numerical simulations of the gravitational waveform from a BBH system consistent with GW150914 were performed. See also Ref.~\cite{Witek:2018dmd} for a focus on the scalar-field dynamics and Ref.~\cite{Shiralilou:2020gah} for recent advancements in the analytical computations of the EsGB Post-Newtonian expansion.
For this theory, one of the best known observational bounds\footnote{Where we adapted the result presented in Ref.~\cite{Yagi:2012gp} to the units of this paper, by comparing the ringdown corrections obtained in this work to the results of Ref.~\cite{Carson:2020cqb}, obtaining a conversion factor of $\sqrt[4]{16\pi \cdot Y}$, where $Y=2-10$ due to the uncertainty in the parameters reconstruction. For $Y \sim 10$ our units are consistent with those of Ref.~\cite{Witek:2018dmd}. Lacking an explicit prediction, for simplicity we also apply the same $Y$ conversion factor to the dCS bound.} is $\ell \lesssim 6-10 \, \mathrm{km}$, estimated from X-ray observations of low-mass BHs in Ref.~\cite{Yagi:2012gp} (although it has to be noted such systems are much more prone to astrophysical uncertainties with respect to GW observations). A comparable bound was derived in Ref.~\cite{Nair:2019iur, Tahura:2019dgr} by applying to this theory the constraints obtained by the LVC from the early inspiral phase of a few GWTC-1 events \cite{LIGOScientific:2019fpa}. As noted in Refs.\cite{Witek:2018dmd,Maselli:2014fca} a theoretical bound from the existence of a BH in EsGB gravity imposes an upper limit\footnote{This is actually a conservative upper limit, since depending on the BH spin and on the specific coupling with the scalar-field the bound decreases.} on the coupling $\frac{\ell \,c^2}{G\,M} < \sqrt{0.691}$. Consequently, from the lightest BH observed, Ref.~\cite{Witek:2018dmd} derived an upper bound $\ell < 6.6 \, \mathrm{km}$. A similar bound ($\ell < 5.4 \, \mathrm{km}$) can be obtained by considering neutron stars in this theory~\cite{Witek:2018dmd, Pani:2011xm}, although in this case the result depends on the equation of state.
Here we note, for the first time to the best of our knowledge, that a tighter bound can be obtained by interpreting the secondary object of the GW190814 binary~\cite{Abbott:2020khf} as a BH. This binary hosted a secondary component with a mass $M = 2.5-2.67 \, M_{\odot}$, either the lightest black hole or the heaviest neutron star ever discovered in a double compact-object system\footnote{Given the large mass ratio of the binary, tidal effects are highly suppressed and consequently these two hypotheses could not be distinguished.}. If the secondary object is indeed interpreted as a BH, this observation provides the bound $\ell < 3.3 \, \mathrm{km}$, which is the \textit{tightest} ever obtained on this theory. The interpretation of the GW190814 secondary object as a BH is supported by various studies of neutron stars maximum mass~\cite{Essick:2020ghc, Tews:2020ylw, Nathanail:2021tay}.\\
Instead, for dCS gravity, combining NICER~\cite{Riley_2019, Miller_2019, inproceedings} and LIGO-Virgo neutron stars observations~\cite{PhysRevLett.119.161101}, Ref.~\cite{Silva:2020acr} recently put forward the most stringent estimate to date on the dCS length scale of $\ell \lesssim 30-50 \, \mathrm{km}$. No constraints from the inspiral of GWTC-1 events was possible in this case, due to the failure of the small-coupling assumption for the parameters of these events~\cite{Nair:2019iur}. See also Ref.~\cite{Okounkova:2019zjf} for the prospects of constraining dCS using numerical simulations of BBH mergers and LIGO-Virgo observations and Ref.~\cite{1843009} for an analysis using the non-detection of amplitude birifringence in GWTC-2 signals to obtain a bound on non-dynamical Chern-Simons $\ell \lesssim 10^3 \, \mathrm{km}$. Both EsGB and dCS gravity evade constraints from binary-pulsar observations~\cite{Nair:2019iur}; solar-system measurements~\cite{Will:2014kxa} can only place bounds on dCS several orders of magnitudes weaker than those presented here, while no bound on EsGB gravity can be obtained from these latter observations~\cite{Nair:2019iur}. It should be stressed that all the aforementioned bounds rely on measurements obtained in the non-dynamical or low-spin regime, while measurements extracted from the merger-ringdown regimes of binary black coalescences, discussed in this manuscript, are obtained from a highly-dynamical, highly spinning remnant compact object dynamics.\\
\textit{\textbf{Dimensionful coupling:} $\mathbf{p=6}$ --} For this case, the obtained bound is $\ell \lesssim 42 \, \mathrm{km}$.
The class of EFTs constructed in Ref.~\cite{Endlich:2017tqa} is one of the most important examples of theories whose spectra possess a $p=6$ correction.
Comparing with previous analyses, this result is both consistent with what predicted by the QNM analysis of Ref.~\cite{PhysRevLett.121.251105} (for a generic formalism including this class of EFTs, see Ref.~\cite{McManus:2019ulj}), and improves on the results\footnote{From the results of Ref.~\cite{PhysRevLett.121.251105} it can be deduced that the units in which this results is quoted are the same as the ones used in this paper up to factors of $O(1)$.} presented in Ref.~\cite{Sennett:2019bpc}. This latter work analysed the early-inspiral phases of a few GWTC-1 observations, disfavouring the appearance of new physics on a scale larger than $\sim 150 \,  \mathrm{km}$.\\
In the dimensionful cases, the measurement of the $\ell$ parameter is still strongly affected by the low number of events considered. Louder and more numerous detections from future gravitational-waves detector network observing runs will significantly shrink the bounds presented. Improvements in the high-frequency sensitivity band of GW detectors will have the highest impact on this measurement, since being able to probe merger-ringdown emission from low-mass BHs will translate into sensitivity on smaller scales (higher energy) effects.
The bounds presented on the Schwarzschild contribution to the spectrum open up the possibility of constraining those theories for which only spherically symmetric solutions are known, but the rotating case remains out of grasp (e.g. Refs.~\cite{Cardoso:2019mqo,PhysRevLett.121.251105,Suvorov:2021amy}). For future prospects on constraining modified theories of gravity with existing and planned GW detectors see Refs.~\cite{Chamberlain:2017fjl, Carson:2019kkh, Perkins:2020tra}. It has to be remarked once again that almost all the constraints discussed, both the ones obtained in this work and those quoted from the literature, were obtained at linear order in the beyond-GR coupling. Given the lack of large observable deviations from the predictions of GR even in the most violent stages of BBH coalescences, this appears to be a justified assumption.
Nonetheless, confirmation of this assumption in a complete non-perturbative treatment at the theoretical level remains of paramount importance; in absence of such a treatment the constraints presented can still be reliably interpreted by treating the beyond-GR theory considered as an effective field-theory.
The only exception to this perturbative approximation is the bound on EsGB gravity, valid in the exact theory, obtained when considering the GW190814 companion as a light BH.
In summary, the results obtained are either superior to previously known bounds or will soon be competitive with current bounds, leveraging on the large number of expected detections and increase in loudness of signals registered by the interferometric gravitational-wave detectors network.

\section{Conclusions}
\label{sec:Conclusions}
We presented the first application of a high-spin version of the ParSpec formalism to observational data collected from the LIGO-Virgo interferometers during the O1-O2-O3a observing runs. Theoretical frameworks as the one considered in this work, enable to extract tighter constraints on modified-gravity deviation parameters with respect to less stringent parametrizations.
We showed how combining current detections within ParSpec allows to start placing bounds on BH spectra at the $\mathcal{O}(5-10\%)$ level for theories including scalar couplings in the action. The length scale of new physics induced by modified theories of gravity including dimensionful couplings is constrained to be $\ell \lesssim 20-40 \, \mathrm{km}$, depending on the coupling mass dimension. The bounds obtained either exceed previous ones present in the literature, or are comparable to some of the best-known estimates derived from other experiments, depending on the specific theory considered. Unlike previous results, such bounds were obtained in a highly dynamical, highly spinning regime. Upcoming data releases and future observing runs from the LIGO-Virgo-Kagra network promise steady improvements to the measurements presented. The measurement precision may soon enter the regime where corrections from Effective Field Theories of beyond-GR gravity could be detectable, offering the concrete prospect to detect modified gravity signatures in BH spacetimes or, alternatively, constrain the fundamental coupling constants of beyond-GR theories to unprecedented precision.\\
Possible extensions of this work include employing an underlying GR template in which the complex amplitudes are constrained as functions of astrophysical parameters, while still allowing to fit the whole post-merger signal, as in Refs.~\cite{Forteza:2020hbw, Damour_Nagar_ring}. Such a template would decrease the number of free parameters and improve on the constraints presented. Another possibility would be to consider different parametrizations, either using a non-polynomial form of QNM parameters or directly casting constraint on an effective metric encompassing a wide range of beyond-GR BH solutions (possibly employing an eikonal approximation). An application of this formalism to specific modified theories of gravity where explicit predictions for the spectral deviation parameters are available is already ongoing.
Finally, it would be interesting to consider an extension of this formalism considering $p$ as a free parameter (dropping the integer assumption), going beyond the current class of theories investigated. Such an addition will likely introduce non-trivial correlations in the parameter space and its feasibility should carefully be explored.

\begin{acknowledgments}

\textit{Software} 
The \texttt{pyRing} package is publicly available at: \href{https://git.ligo.org/lscsoft/pyring}{https://git.ligo.org/lscsoft/pyring},
where example configuration files using the ParSpec framework are also provided.
LIGO-Virgo data are interfaced through \texttt{GWpy}~\cite{gwpy} and some of the post-processing steps are handled through \texttt{PESummary}~\cite{pesummary}, a sampler-agnostic \texttt{python} package providing interactive webpages to simplify results visualisation. Moreover, \texttt{PESummary} meta-files are used to store the complete information (both of the internal \texttt{pyRing} parameters and of the software environment) for the analysis to be completely reproducible. 
Other open-software \texttt{python} packages, accessible through \texttt{PyPi}, internally used by \texttt{pyRing} are: \texttt{corner, gwsurrogate, matplotlib, numba, numpy, seaborn, surfinBH}~\cite{corner, gwsurrogate, matplotlib, numba, numpy, seaborn, surfinbh1, surfinbh2}.

\textit{Acknowledgments}
The author would like to thank Alessandra Buonanno, Nathan Johnson-McDaniel and Bangalore Sathyaprakash for useful discussions, Emanuele Berti, Walter Del Pozzo, Danny Laghi, Andrea Maselli and Paolo Pani for helpful comments on the manuscript. I would also like to thank the anonymous referees who helped to significantly improve the readability of the manuscript. This work greatly benefited from discussions within the \textit{Testing General Relativity} working group of the LIGO-Virgo-Kagra collaboration. This research has made use of data, software and/or web tools obtained from the Gravitational Wave Open Science Center (https://www.gw-openscience.org), a service of LIGO Laboratory, the LIGO Scientific Collaboration and the Virgo Collaboration. LIGO is funded by the U.S. National Science Foundation. Virgo is funded by the French Centre National de Recherche Scientifique (CNRS), the Italian Istituto Nazionale della Fisica Nucleare (INFN) and the Dutch Nikhef, with contributions by Polish and Hungarian institutes.\\
\textit{We would like to thank all of the essential workers who put their health at risk during the COVID-19 pandemic, without whom we would not have been able to complete this work.}
\end{acknowledgments}

\begin{table*}[t]
\caption{Median and $90 \%$ credible levels of ringdown parameters deviation distributions; $p$ indicates the coupling dimension, $j$ the index of the expansion order, while $\mathrm{M}_{max}$ the maximum expansion order considered, as defined in Eq.~(\ref{eq:ParSpec_expansion}). No statistically significant deviation from the predictions of BH dynamics in GR is found.\\}

\begin{ruledtabular}
\begin{tabular}{ccccc}
$(j, \mathrm{M}_{max}$) & \hspace{0.15cm}$\delta\omega_{220}$ & \hspace{0.15cm}$\delta\omega_{221}$ & \hspace{0.15cm}$\delta\tau_{220}$ & \hspace{0.15cm}$\delta\tau_{221}$ \vspace{0.1cm} \\ 
\hline
\hline
& &\, $p=0$ & & \vspace{0.08cm} \\
\hline
\hline
\, (0,0) & ${-0.05}^{+0.05}_{-0.05}$ & ${-0.04}^{+0.18}_{-0.15}$ & \hspace{0.14cm} ${0.01}^{+0.22}_{-0.17}$ & \hspace{0.14cm} ${0.07}^{+0.31}_{-0.33}$ \\ \vspace{0.05cm}
(0,1) & ${-0.06}^{+0.09}_{-0.09}$ & ${-0.03}^{+0.19}_{-0.17}$ & \hspace{0.14cm} ${0.02}^{+0.22}_{-0.18}$ & \hspace{0.14cm} ${0.08}^{+0.30}_{-0.35}$ \\ \vspace{0.05cm}
(0,2) & ${-0.00}^{+0.33}_{-0.32}$ & ${-0.05}^{+0.35}_{-0.32}$ & \hspace{0.14cm} ${0.02}^{+0.21}_{-0.18}$ & \hspace{0.14cm} ${0.22}^{+0.20}_{-0.41}$ \\ \vspace{0.05cm}
(1,1) & ${-0.07}^{+0.36}_{-0.31}$ & ${-0.01}^{+0.34}_{-0.35}$ & \hspace{0.14cm} ${0.08}^{+0.31}_{-0.45}$ & \hspace{0.14cm} ${0.01}^{+0.38}_{-0.39}$ \\ \vspace{0.05cm}
(1,2) & \hspace{0.14cm} ${0.01}^{+0.35}_{-0.37}$ & ${-0.02}^{+0.36}_{-0.33}$ & ${-0.02}^{+0.39}_{-0.36}$ & ${-0.01}^{+0.38}_{-0.36}$ \\ \vspace{0.05cm}
(2,2) & \hspace{0.14cm} ${0.04}^{+0.11}_{-0.12}$ & \hspace{0.14cm} ${0.02}^{+0.13}_{-0.13}$ & \hspace{0.14cm} ${0.01}^{+0.37}_{-0.38}$ & \hspace{0.14cm} ${0.04}^{+0.17}_{-0.18}$ \\
\hline
\hline
& &\, $p=2$ & & \vspace{0.08cm} \\
\hline
\hline
\, (0,0) & ${-0.10}^{+0.14}_{-0.16}$ & ${-0.04}^{+0.37}_{-0.33}$ & ${-0.04}^{+0.41}_{-0.31}$ & \hspace{0.14cm} ${0.01}^{+0.37}_{-0.40}$ \\ \vspace{0.05cm}
(0,1) & ${-0.12}^{+0.18}_{-0.17}$ & ${-0.03}^{+0.38}_{-0.32}$ & ${-0.04}^{+0.39}_{-0.33}$ & ${-0.02}^{+0.40}_{-0.37}$ \\ \vspace{0.05cm}
(0,2) & ${-0.17}^{+0.36}_{-0.22}$ & ${-0.01}^{+0.36}_{-0.35}$ & \hspace{0.14cm} ${0.01}^{+0.33}_{-0.32}$ & \hspace{0.14cm} ${0.00}^{+0.36}_{-0.38}$ \\ \vspace{0.05cm}
(1,1) & ${-0.17}^{+0.44}_{-0.23}$ & \hspace{0.14cm} ${0.01}^{+0.36}_{-0.38}$ & ${-0.02}^{+0.39}_{-0.37}$ & \hspace{0.14cm} ${0.07}^{+0.33}_{-0.44}$ \\ \vspace{0.05cm}
(1,2) & ${-0.04}^{+0.36}_{-0.34}$ & ${-0.02}^{+0.39}_{-0.37}$ & \hspace{0.14cm} ${0.02}^{+0.35}_{-0.41}$ & ${-0.02}^{+0.40}_{-0.37}$ \\ \vspace{0.05cm}
(2,2) & \hspace{0.14cm} ${0.09}^{+0.18}_{-0.17}$ & \hspace{0.14cm} ${0.07}^{+0.28}_{-0.33}$ & ${-0.02}^{+0.39}_{-0.36}$ & \hspace{0.14cm} ${0.05}^{+0.34}_{-0.37}$ \\
\hline
\hline
& &\, $p=4$ & & \vspace{0.08cm} \\
\hline
\hline
\, (0,0) & ${-0.10}^{+0.20}_{-0.26}$ & \hspace{0.14cm} ${0.06}^{+0.29}_{-0.42}$ & ${-0.06}^{+0.39}_{-0.31}$ & \hspace{0.14cm} ${0.01}^{+0.37}_{-0.38}$ \\ \vspace{0.05cm}
(0,1) & ${-0.13}^{+0.27}_{-0.22}$ & \hspace{0.14cm} ${0.03}^{+0.34}_{-0.39}$ & ${-0.05}^{+0.39}_{-0.30}$ & \hspace{0.14cm} ${0.03}^{+0.35}_{-0.40}$ \\ \vspace{0.05cm}
(0,2) & ${-0.07}^{+0.38}_{-0.31}$ & ${-0.08}^{+0.41}_{-0.30}$ & ${-0.07}^{+0.42}_{-0.30}$ & ${-0.04}^{+0.41}_{-0.35}$ \\ \vspace{0.05cm}
(1,1) & ${-0.11}^{+0.42}_{-0.29}$ & ${-0.03}^{+0.41}_{-0.35}$ & ${-0.01}^{+0.39}_{-0.39}$ & \hspace{0.14cm} ${0.05}^{+0.35}_{-0.44}$ \\ \vspace{0.05cm}
(1,2) & ${-0.02}^{+0.38}_{-0.37}$ & \hspace{0.14cm} ${0.04}^{+0.33}_{-0.41}$ & ${-0.04}^{+0.42}_{-0.36}$ & ${-0.02}^{+0.40}_{-0.36}$ \\ \vspace{0.05cm}
(2,2) & \hspace{0.14cm} ${0.16}^{+0.21}_{-0.28}$ & \hspace{0.14cm} ${0.03}^{+0.32}_{-0.36}$ & \hspace{0.14cm} ${0.01}^{+0.38}_{-0.40}$ & \hspace{0.14cm} ${0.01}^{+0.35}_{-0.37}$ \\
\hline
\hline
& &\, $p=6$ & & \vspace{0.08cm} \\
\hline
\hline
\, (0,0) & ${-0.12}^{+0.30}_{-0.25}$ & ${-0.01}^{+0.37}_{-0.38}$ & ${-0.00}^{+0.36}_{-0.35}$ & \hspace{0.14cm} ${0.05}^{+0.33}_{-0.40}$ \\ \vspace{0.05cm}
(0,1) & ${-0.10}^{+0.33}_{-0.26}$ & ${-0.03}^{+0.37}_{-0.36}$ & ${-0.02}^{+0.39}_{-0.35}$ & ${-0.04}^{+0.39}_{-0.34}$ \\ \vspace{0.05cm}
(0,2) & ${-0.07}^{+0.37}_{-0.31}$ & ${-0.02}^{+0.39}_{-0.35}$ & ${-0.04}^{+0.39}_{-0.33}$ & \hspace{0.14cm} ${0.06}^{+0.33}_{-0.44}$ \\ \vspace{0.05cm}
(1,1) & ${-0.08}^{+0.43}_{-0.32}$ & ${-0.01}^{+0.38}_{-0.36}$ & ${-0.01}^{+0.39}_{-0.38}$ & \hspace{0.14cm} ${0.02}^{+0.37}_{-0.40}$ \\ \vspace{0.05cm}
(1,2) & ${-0.09}^{+0.45}_{-0.30}$ & \hspace{0.14cm} ${0.04}^{+0.35}_{-0.42}$ & ${-0.05}^{+0.43}_{-0.34}$ & ${-0.04}^{+0.39}_{-0.34}$ \\ \vspace{0.05cm}
(2,2) & \hspace{0.14cm} ${0.13}^{+0.25}_{-0.34}$ & \hspace{0.14cm} ${0.04}^{+0.32}_{-0.36}$ & ${-0.01}^{+0.39}_{-0.37}$ & \hspace{0.14cm} ${0.02}^{+0.36}_{-0.36}$ \\
\end{tabular}
\end{ruledtabular}
\label{tab:bounds_table}
\end{table*}

\begin{table*}[t]
\caption{Natural logarithm of the Bayes Factors comparing the hypotheses that all the BH spectra observed by LIGO-Virgo are described by BH dynamics in GR against an alternative dynamics; $p$ indicates the coupling parameter dimension, $\mathrm{M}_{max}$ the maximum expansion order considered, both defined in Eq.~(\ref{eq:ParSpec_expansion}). The error on single-events Bayes factor is $\pm 0.1$, the error on the combined Bayes factor is $\pm 0.6$. A large number of cases yield significantly large negative values, indicating that, with available data, the GR predictions are highly favoured for these parameters. In no case evidence for a statistically significant deviation is found.\\}

\begin{ruledtabular}
\begin{tabular}{ccccc}
$\mathrm{M}_{max}$ & \hspace{0.15cm}$\delta\omega_{220}$ & \hspace{0.15cm}$\delta\omega_{221}$ & \hspace{0.15cm}$\delta\tau_{220}$ & \hspace{0.15cm}$\delta\tau_{221}$ \vspace{0.1cm} \\ 
\hline
\hline
& &\, $p=0$ & & \vspace{0.08cm} \\
\hline
\hline
\, 0 & ${-14.55}$ & ${-2.25}$ & ${-1.23}$ & ${-0.19}$ \\ \vspace{0.05cm}
1 & ${-16.66}$ & ${-4.35}$ & ${-2.86}$ & ${-1.01}$ \\ \vspace{0.05cm}
2 & ${-26.28}$ & ${-8.66}$ & ${-3.52}$ & ${-2.22}$ \\
\hline
\hline
& &\, $p=2$ & & \vspace{0.08cm} \\
\hline
\hline \, 0 & ${-3.11}$ & ${-1.49}$ & ${-1.40}$ & ${-0.02}$ \\ \vspace{0.05cm}
1 & ${-3.18}$ & ${-1.00}$ & ${-1.86}$ & ${-1.16}$ \\ \vspace{0.05cm}
2 & ${-6.88}$ & ${-3.32}$ & ${-2.45}$ & ${-1.02}$ \\
\hline
\hline
& &\, $p=4$ & & \vspace{0.08cm} \\
\hline
\hline
\, 0 & ${-1.21}$ & ${-0.74}$ & ${-1.30}$ & ${-0.51}$ \\ \vspace{0.05cm}
1 & ${-2.25}$ & ${-1.13}$ & ${-2.19}$ & ${-1.12}$ \\ \vspace{0.05cm}
2 & ${-4.23}$ & ${-3.91}$ & ${-2.64}$ & ${-1.27}$ \\
\hline
\hline
& &\, $p=6$ & & \vspace{0.08cm} \\
\hline
\hline
\, 0 & ${-1.37}$ & ${-1.40}$ & ${-1.68}$ & ${-0.61}$ \\ \vspace{0.05cm}
1 & ${-1.87}$ & ${-1.29}$ & ${-2.09}$ & ${-0.57}$ \\ \vspace{0.05cm}
2 & ${-5.36}$ & ${-1.07}$ & ${-2.46}$ & ${-0.87}$ \\
\end{tabular}
\end{ruledtabular}
\label{tab:logBs}
\end{table*}

\begin{table*}[t]
\caption{Upper bounds at the $90 \%$ credible level on $\ell$ in km units, when considering deviations in different ringdown parameters; $p$ indicates the coupling parameter dimension, $\mathrm{M}_{max}$ the maximum expansion order considered, both defined in Eq.~(\ref{eq:ParSpec_expansion}). For p=0 the coupling is reabsorbed in the deviation parameter. The bounds presented are either competitive or superior, depending on the theory considered, with those obtained through other experiments and improve upon previous bounds obtained with GW observations.\\}

\begin{ruledtabular}
\begin{tabular}{ccccc}
$\mathrm{M}_{max}$ & \hspace{0.15cm}$\delta\omega_{220}$ & \hspace{0.15cm}$\delta\omega_{221}$ & \hspace{0.15cm}$\delta\tau_{220}$ & \hspace{0.15cm}$\delta\tau_{221}$ \vspace{0.1cm} \\ 
\hline
\hline
& &\, $p=2$ & & \vspace{0.08cm} \\
\hline
\hline
\, 0 & ${43.04}$ & ${46.02}$ & ${57.46}$ & ${62.96}$ \\ \vspace{0.05cm}
1 & ${31.57}$ & ${62.88}$ & ${59.73}$ & ${48.47}$ \\ \vspace{0.05cm}
2 & ${23.35}$ & ${44.28}$ & ${60.02}$ & ${52.07}$ \\
\hline
\hline
& &\, $p=4$ & & \vspace{0.08cm} \\
\hline
\hline
\, 0 & ${45.35}$ & ${59.52}$ & ${54.30}$ & ${43.01}$ \\ \vspace{0.05cm}
1 & ${48.28}$ & ${63.93}$ & ${56.54}$ & ${62.99}$ \\ \vspace{0.05cm}
2 & ${35.35}$ & ${60.60}$ & ${54.23}$ & ${53.74}$ \\
\hline
\hline
& &\, $p=6$ & & \vspace{0.08cm} \\
\hline
\hline
\, 0 & ${47.78}$ & ${55.88}$ & ${52.40}$ & ${54.97}$ \\ \vspace{0.05cm}
1 & ${36.57}$ & ${59.71}$ & ${46.97}$ & ${56.57}$ \\ \vspace{0.05cm}
2 & ${41.76}$ & ${61.51}$ & ${57.80}$ & ${44.45}$ \\
\end{tabular}
\end{ruledtabular}
\label{tab:ell_eff_table}
\end{table*}

\clearpage

\bibliographystyle{apsrev}
\bibliography{References}

\end{document}